\documentclass[pdflatex,sn-mathphys-num]{sn-jnl}% Math and Physical Sciences Numbered Reference Style 
\usepackage{graphicx}%
\usepackage{multirow}%
\usepackage{amsmath,amssymb,amsfonts}%
\usepackage{amsthm}%
\usepackage{mathrsfs}%
\usepackage[title]{appendix}%
\usepackage{xcolor}%
\usepackage{textcomp}%
\usepackage{manyfoot}%
\usepackage{booktabs}%
\usepackage{algorithm}%
\usepackage{algorithmicx}%
\usepackage{algpseudocode}%
\usepackage{listings}%

\usepackage{amssymb}
\usepackage{subcaption}
\usepackage{multirow}
\usepackage{longtable}
\usepackage{amsmath}
\usepackage{booktabs}
\graphicspath{ {./figs/} }
\usepackage{float}
\usepackage{enumitem}
\usepackage{appendix}

\theoremstyle{thmstyleone}%
\theoremstyle{thmstyletwo}%

\theoremstyle{thmstylethree}%

\begin{document}

\title[Article Title]{From Pro, Anti to Informative and Hesitant: An Infoveillance study of COVID-19 vaccines and vaccination discourse on Twitter}


\author*[1]{\fnm{Pardeep} \sur{Singh}}\email{pardee87\_scs@jnu.ac.in}

\author[2]{\fnm{Rabindra} \sur{Lamsal}}\email{rlamsal@student.unimelb.edu.au}

\author[1]{\fnm{Monika} \sur{Singh}}\email{monika095\_scs@jnu.ac.in}

\author[1]{\fnm{Satish} \sur{Chand}}\email{schand@mail.jnu.ac.in}

\author[1]{\fnm{Bhawna} \sur{Shishodia}}\email{bhawna.shishodia.du@gmail.com}

\affil*[1]{\orgdiv{School of Computer \& Systems Sciences}, \orgname{Jawaharlal Nehru University}, \orgaddress{\state{New Delhi},\postcode{110067}, \country{India}}}

\affil[2]{\orgdiv{School of Computing and Information Systems}, \orgname{The University of Melbourne}, \orgaddress{  \state{Victoria}, \postcode{3010}, \country{}Australia}}


\abstract{COVID-19 pandemic has brought unprecedented challenges to the world, and vaccination has been a key strategy to combat the disease. Since Twitter is one of the most widely used public microblogging platforms, researchers have analysed COVID-19 vaccines and vaccination Twitter discourse to explore the conversational dynamics around the topic. While contributing to the crisis informatics literature, we curate a large-scale geotagged Twitter dataset, \textit{GeoCovaxTweets Extended}, and explore the discourse through multiple spatiotemporal analyses. This dataset covers a longer time span of 38 months, from the announcement of the first vaccine to the availability of booster doses. Results show that 43.4\% of the collected tweets, although containing phrases and keywords related to vaccines and vaccinations, were unrelated to the COVID-19 context. In total, 23.1\% of the discussions on vaccines and vaccinations were classified as Pro, 16\% as Hesitant, 11.4\% as Anti, and 6.1\% as Informative. The trend shifted towards Pro and Informative tweets globally as vaccination programs progressed, indicating a change in the public's perception of COVID-19 vaccines and vaccination. Furthermore, we explored the discourse based on account attributes, i.e., followers' counts, and tweet counts. Results show a significant pattern of discourse differences. Our findings highlight the potential of harnessing a large-scale geotagged Twitter dataset to understand global public health communication and to inform targeted interventions aimed at addressing vaccine hesitancy.}


\maketitle

\section{Introduction}\label{sec1}

COVID-19, caused by the Severe Acute Respiratory Syndrome Coronavirus 2 (SARS-CoV-2), has claimed the lives of above 6 million people globally between January 1, 2020, and December 31, 2022 \citep{taylor2022covid}. On March 11, 2020, the World Health Organization (WHO) declared the COVID-19 outbreak a pandemic as the virus rapidly spread across the globe, causing widespread illness and death. Social distancing, face masks, and hand sanitation were advised to reduce the spread of the virus. Governments worldwide temporarily closed educational institutions and suspended international travel while outlawing large gatherings. The impacts of these measures were seen on the economy, and mental health \citep{usher2020covid}. As a result, vaccine development became a major focus of the scientific and medical communities. The global effort to develop effective vaccines against COVID-19 did bring promising results. As of April 8, 2022, the WHO has evaluated 10 different COVID-19 vaccines to have met the necessary criteria for safety and efficacy\footnote{https://www.who.int/emergencies/diseases/novel-coronavirus-2019/covid-19-vaccines/advice}.

As the vaccine rollout started worldwide in different phases, opinions on the vaccine's safety and efficacy surfaced on social media platforms such as Twitter. Some users expressed support for the vaccines, citing their effectiveness in preventing the spread of the virus, while others expressed doubt, citing concerns about the speed of the vaccine distribution and potential side effects \citep{gisondi2022deadly,muric2021covid}. Still, others expressed uncertainty, noting that the decision to get vaccinated is personal. Most importantly, hesitancy in getting vaccinated for COVID-19 has been cited for various factors, such as misinformation, lack of trust in the medical system, or concerns about the safety and effectiveness of vaccines \citep{lazarus2022revisiting}. Anti-vaccination groups had been observed using false claims of a causal link between vaccines and unrelated illnesses as a tactic \footnote{https://www.bbc.com/future/article/20211209-how-to-talk-to-vaccine-hesitant-people}. However, these claims were based on misinterpreting correlation and causation. Correlation means that two events occurred together but does not imply that one event caused the other. Just because someone got sick after vaccination does not mean the vaccine caused the illness. Therefore, it was extremely necessary for researchers in crisis informatics and social computing to explore the nature of these misinformation sources, the respective actors in the information propagation network, and the conversational dynamics of the contexts around vaccines and vaccination. 

With its large and active user base, Twitter has been extensively explored for discourse analysis related to COVID-19. Its live and historical public feed can be accessed through various API endpoints, making it a valuable source for natural language processing (NLP) tasks. Researchers have utilized Twitter discussions to analyze public opinions and attitudes towards COVID-19 vaccines, and measure vaccine hesitancy \citep{cotfas2021longest,delcea2022new,kwok2021tweet,loomba2021measuring,deverna2021covaxxy,muric2021covid}. However, previous studies are limited in their regional focus and temporal coverage, highlighting the need for a more comprehensive global analysis of the discourse surrounding COVID-19 vaccines on Twitter over a longer period. Additionally, those studies classified the discourse into simplistic pro-anti-neutral or pro-anti-neutral-hesitant classifications, but a more nuanced understanding of public perceptions and attitudes towards COVID-19 vaccines is necessary. 

\subsection{Research Objectives}

As a response to the ongoing COVID-19 pandemic and the desire to gain a better comprehension of the global discussion around COVID-19 vaccines and vaccination, a large-scale geotagged Twitter dataset --- \textit{GeoCovaxTweets Extended} --- is created. The dataset covers a 38-month period, including the first COVID-19 vaccine announcement to the availability of booster doses, and contains tweets from around the globe. We examine the dataset to attain a more comprehensive understanding of the global discourse about COVID-19 vaccines and vaccination, across the following distinct perspectives: support, criticism, hesitancy, information, and irrelevant. Our study highlights the spatial-temporal evolution of the discourse on Twitter, providing beneficial insights for policymakers and public health officials. Specifically, we look at the following research questions:
\begin{itemize}
    \item \textbf{RQ1}: How has the rollout of COVID-19 vaccines affected social media discourse, particularly on Twitter?
    
    \item \textbf{RQ2}: How do Twitter users from different countries and territories participate in the discourse surrounding COVID-19 vaccines and vaccination, and how does this participation vary depending on their stance (pro, anti, hesitant, informative)?

    \item \textbf{RQ3}: How do verified Twitter accounts shape the discourse on COVID-19 vaccines and vaccination, and what messages do they promote?

    \item \textbf{RQ4}: How has the discourse around COVID-19 vaccines and vaccination changed over time, and how does this relate to the global vaccination progress?

    \item \textbf{RQ5}: Is there any usage of contextual hashtags in tweets related to COVID-19 vaccines and vaccination, and if so, how do they vary across different stances? 
    
    \item \textbf{RQ6}: How does the level of Twitter user engagement, measured by followers and tweet count, vary across different stances and over time in the discourse surrounding COVID-19 vaccines and vaccination?
     
\end{itemize}
While addressing the above research questions, this paper contributes the following to the crisis informatics literature (\textbf{data releases will be done after peer review completes}):
\begin{itemize}   
\item Most existing studies analyzing the COVID-19 vaccines and vaccination discourse are categorically limited. Therefore, we explore the Twitterverse while considering the following classes: Pro, anti, informative, hesitant, and not-related, for an extended period of over 165 weeks. We investigate how Twitter users globally and from top countries in the discourse, and distinct stances engage in vaccine discourse, analyze the role of verified Twitter accounts, track changes in vaccine discourse over time and relate it to global vaccination progress, study the use of contextual hashtags, and measure Twitter engagement levels across different stances and time periods. 

\item We release a new labelled dataset --- \textit{PAIHcovax} --- for studying COVID-19 vaccines and vaccination stances. We experiment with multiple state-of-the-art transformer-based models on the dataset and release the best-performing classifier~---~\textit{CovaxBERT}.

\item Furthermore, we release a large-scale COVID-19 vaccines and vaccination-specific tweets dataset, \textit{GeoCovaxTweets Extended}, to facilitate further research on this topic. As per Twitter policy, we only share tweet identifiers with the public. Researchers need to retrieve the original tweets and the related metadata with the help of libraries/applications such as \textit{twarc} (Python library) and \textit{Hydrator} (Desktop application).

\end {itemize}

\section{Related work} 

Numerous publicly available datasets comprising massive collections of tweets related to the COVID-19 pandemic have been made accessible \citep{chen2020tracking, banda2021large, lamsal2021design, imran2022tbcov, lamsal2023billioncov}. However, these large-scale datasets mainly focus on general COVID-19 discussions. Researchers have also developed publicly available tweet datasets that cover vaccines and anti-vaccine discourse to better understand the prevalence and sentiment of vaccine hesitancy on social media. Some of these datasets are language-specific, primarily in English \citep{deverna2021covaxxy, muric2021covid}, while others are multilingual \citep{chen2022multilingual, di2022vaccineu} and region-specific \citep{hu2021revealing, kwok2021tweet, chen2022multilingual}. These datasets are a valuable resource for researchers, public health officials, and policymakers seeking to better comprehend this phenomenon. For instance, \citep{deverna2021covaxxy} presented a dataset of 4.7 million English language vaccine-related tweets extracted using 76 hashtags and keywords for the timeline January 4--11, 2021. Similarly, \citep{muric2021covid} introduced a dataset of 137 million tweets containing anti-vaccine conversations on Twitter. The dataset comprised tweets collected through the streaming API using 60 keywords and historical tweets of 70,000 users propagating anti-vaccine tweets. \citep{hu2021revealing} presented a dataset of 0.3 million geotagged tweets, focusing on the United States, collected between March 01, 2020, and February 08, 2021. Additionally, \citep{kwok2021tweet} collected a dataset of 31.1k geotagged tweets specific to Australia from January 20, 2020, to October 20, 2022. Social science researchers have also used survey-based methods, such as self-reporting questionnaires, to analyze people's stance toward vaccines and vaccination on a limited scale \citep{sallam2021covid,bonnevie2021quantifying}.

Some studies have focused on sentiment or stance analysis of the COVID-19 vaccines and vaccination discourse. Researchers have used sentiment analyzers such as TextBlob, VADER or supervised classifiers to perform sentiment analysis \citep{khan2021us,luo2022understanding,qorib2023covid}.
\citep{chandrasekaran2022examining} examined public sentiments and attitudes towards COVID-19 vaccination using Twitter data. The authors analyzed 2.94 million COVID-19 vaccine-related tweets created between January 1, 2021, and April 31, 2021. The authors performed topic modeling to identify underlying topics and used VADER to measure sentiment scores. However, sentiment analyzers such as TextBlob and VADER rely heavily on pre-defined dictionaries and sentiment lexicons, which may not always capture the nuances of language. Hence, another option is to label datasets manually with their respective classes (e.g., pro, anti, neutral) and use supervised classifiers for the classification task. This allows the model to generalize and make predictions on unseen text data with a higher degree of accuracy.

Continuing with the idea of developing labelled datasets,  a study \citep{cotfas2021longest} collected 2.3 million English language tweets related to the COVID-19 vaccine during the first month following the announcement of the vaccines in the United Kingdom from November 2020 to December 2020. The authors presented 7,530 labelled tweets and examined the sentiment while categorizing them into three classes: favour, against, and neutral. \citep{poddar2022winds} presented a vaccine-related dataset covering the timeline from January 2018 to March 2021 and divided the tweet collection timeline into three periods: pre-COVID (January 2018 to December 2019), COVID-19 period (2020), and post-COVID-19 period (January-March 2021). The authors developed a labelled dataset of 1700 tweets covering a 14 months timeline, January 2020--March 2021, with three classes: anti-vax, pro-vax, and neutral and used BERT and CT-BERT for tweet stance classification. Similarly, \citep{chen2022multilingual} presented a multilingual tweets dataset containing 2.1 million tweets from January 20, 2020, to March 15, 2021, along with 17,934 labelled tweets with five categories: positive, negative, neutral, positive but dissatisfaction, and off-topic. However, the majority of tweets are in the French language. Likewise, \citep{di2022vaccineu} introduced a multilingual dataset containing 3k labelled tweets from November 2020 to November 2021 in French, German and Italian languages with four categories: pro, anti, neutral, and out-of-context. Their focus was on the European context only. \citep{martinez2023spanish} created an annotated corpus of 2,801 Spanish tweets related to COVID-19 vaccination, with three classes --- in favor, against, and neither --- and used a semi-supervised approach to extend the corpus with unlabelled tweets.
\citep{mu2023vaxxhesitancy} presented a dataset of 3,101 tweets in the English language covering 17 months timeline from November 2020--April 2022 while adding a hesitancy class and redefined the COVID-19 vaccine stance classification problem as a four-way classification task: pro, anti, neutral, and hesitant.

Most previous studies \citep{poddar2022winds,cotfas2021longest} have classified vaccine hesitancy into three classes: pro, anti, and neutral, with some introducing the hesitant class \citep{mu2023vaxxhesitancy}. We argue that there is a need for a more comprehensive understanding of online discussions regarding vaccination, particularly in identifying and monitoring concerns expressed by individuals who are hesitant about vaccines. Thus, we introduce three new separate classes, namely, \textit{hesitant}, \textit{informative}, and \textit{not-related}, in addition to \textit{pro} and \textit{anti} for the annotation of utterances in tweet conversations. The \textit{hesitant} class pertains to statements that convey uncertainty or skepticism about vaccines and vaccination, while the \textit{informative} class refers to statements that provide factual information about vaccines and vaccination. The \textit{not-related} class pertains to statements that lack relevance or association with the discourse surrounding COVID-19 vaccines and vaccination, even if they contain relevant keywords or hashtags. Our five-class dataset allows for examining vaccine hesitancy and is essential when analyzing the Twitter discourse around the topic. Table \ref{lit-diff} summarises literature involving labelled datasets similar to \textit{PAIHcovax}. To our knowledge, this study is the first to consider five classes that comprehensively capture the Twitter conversational dynamics of support, criticism, hesitance, information, and irrelevant discourse regarding COVID-19 vaccines and vaccination.

\begin{table}[]
    \caption{Summary of previously published datasets and \textit{PAIHcovax}.}
     
    \label{lit-diff}
    \tiny
    %\centering
\begin{tabular}{l|l|l|l|l} 
\hline
\textbf{Dataset} & \textbf{Timeline}                                                               & \begin{tabular}[c]{@{}l@{}}\textbf{Labelled}\\\textbf{ Tweets}\end{tabular} & \textbf{Labels}                                                                                                                   & \textbf{Language}  \\ 
\hline
\citep{cotfas2021longest}     & \begin{tabular}[c]{@{}l@{}}Nov 2020 -Dec 2020\\ (one month)\end{tabular}    &  2,792                                                                        & \begin{tabular}[c]{@{}l@{}}in favor, against, neutral\end{tabular}                                                              & English                 \\ 
\hline
 \citep{poddar2022winds}  & \begin{tabular}[c]{@{}l@{}}Jan 2020 - March 2021\\ (14 months)\end{tabular} &  1,700                                                                        & \begin{tabular}[c]{@{}l@{}}in favor, against, neutral\end{tabular}                                                              & English                \\ 
\hline
\citep{chen2022multilingual} & \begin{tabular}[c]{@{}l@{}}Jan 2020 -March 2021\\ (14 months)\end{tabular}  &17,934                                                                       & \begin{tabular}[c]{@{}l@{}}positive, negative, neutral,\\positive but dissatisfaction,\\ off-topic\end{tabular}                                                  & French, German, Dutch etc. \\ 
\hline
 \citep{di2022vaccineu}  & \begin{tabular}[c]{@{}l@{}}Nov 2020 - November 2021\\ (12 months)\end{tabular}  &  3,000                                                                        & \begin{tabular}[c]{@{}l@{}}pro, anti, neutral, out-of-context\end{tabular}                                                      & Spanish, German, Italian.        \\ 
\hline
\citep{mu2023vaxxhesitancy} & \begin{tabular}[c]{@{}l@{}}Nov 2020 -April 2022\\ (17 months)\end{tabular}  &  3,101                                                                        & \begin{tabular}[c]{@{}l@{}}pro, anti, hesitancy, irrelevant\end{tabular}                                                       & English                 \\ 
\hline

\citep{martinez2023spanish} & \begin{tabular}[c]{@{}l@{}}Mar 2020 -Jan 2022\\ (22 months)\end{tabular}  &  2,801                                                                        & \begin{tabular}[c]{@{}l@{}}in favor, against, neither\end{tabular}                                                       & Spanish                 \\ 
\hline

\textbf{\textit{PAIHcovax}} (ours)         & \begin{tabular}[c]{@{}l@{}}Jan 2020 - Mar 2023\\ (38 months)\end{tabular}   & 4,950                                                                        & \begin{tabular}[c]{@{}l@{}}Pro-covidvax, Anti-covidvax,\\ Informative-covidvax,\\ Hesitant-covidvax, Not-related\end{tabular} & English                 \\
\hline
\end{tabular}

\end{table}

\section{Method}

\subsection{Data Curation}
\subsubsection{The \textit{GeoCovaxTweets Extended} Dataset}
Researchers use Twitter as their primary data source because it is a popular social media platform with a large user base of 353 million active users. Moreover, Twitter offers a wide range of API endpoints that enable researchers to access publicly available data in a convenient and efficient manner. Some of the widely used API endpoints include \textit{Tweet lookup}, \textit{Search}, \textit{Tweet counts}, and \textit{Filtered stream}. The \textit{Tweet lookup endpoint} allows users to retrieve individual tweets based on their unique identifiers and access information such as likes, retweets, and replies. The \textit{Search endpoint} is another frequently used endpoint that enables researchers to search for tweets containing specific keywords or hashtags within a specific time frame. The \textit{Tweet counts endpoint} allows researchers to estimate the number of tweets that match a particular query, while the \textit{Filtered stream endpoint} provides real-time access to a sample of public tweets that are filtered based on specific keywords or hashtags.

\begin{figure}
\centering
\includegraphics[width=0.6\textwidth]{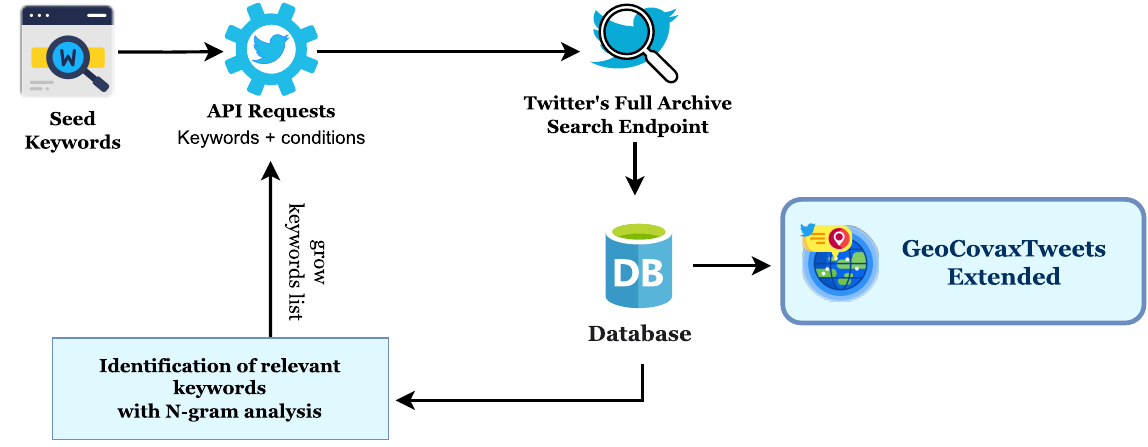}
    \caption{The data curation process.}
    \label{data-curation-process}
\end{figure}

In this study, we utilize Twitter's full-archive search endpoint that returns 100\% of entire Twitter data, unlike filtered and standard search endpoints that return 1\% of Twitter data at a particular instance. To obtain tweets related to the discourse, we implement the sampling technique applied in some of the existing works \citep{deverna2021covaxxy, lamsal2021design}, i.e., starting with a small number of initial tweets and then using those tweets to identify and gather additional tweets that are related to the topic. We started  with  the  following 25 seed  keywords: \textit{\#vaccines, \#vaccination, \#coronavaccine, \#CovidVaccineScam, \#covidvaccinedeaths, \#covidvaccineVictims, \#CovidVaccineKills, \#BanCovidVaccine, \#VaccineFailure,\#vaxxdamage, \#LargestVaccinationDrive, \#ReadyToVaccinate, \#We4Vaccine, \#modernavaccine, \#SputnikVaccine, \#Oxfordvaccine, \#astrazenica, abolishbigpharma, notocoronavirusvaccines, VaccinesAreNotTheAnswer, vaccinesarepoison, vaccinationchoice, VaccineAgenda, Getvaccinated,  NoForcedVaccines, novaccinemandates, saynotovaccines}. The complete list of hashtags and keywords used for tweet collection is shown in Table \ref{hashtags}. We concatenated the collected data with \textit{GeoCovaxTweets} \citep{singh2023geocovaxtweets} to form the final collection~---~\textit{GeoCovaxTweets Extended}. Figure \ref{data-curation-process} illustrates the data curation process. \textit{GeoCovaxTweets Extended} is a multilingual global dataset containing geotagged tweets with a temporal coverage of over 165 weeks, i.e., January 1, 2020--March 05, 2023. We explore this dataset in detail later in Section \ref{exploring-dataset}.
 
\begin{table}[t!]
    \centering
    \caption{
    Hashtags and keywords used for curating \textit{GeoCovaxTweets Extended}.}
      \label{hashtags}
      \footnotesize
    \begin{tabular}{p{\textwidth}}
\hline
antivaccine, antivaxxers, \#BigPharma, bigpharmafia, bigpharmakills, covidvaccineispoison, \#DoctorsSpeakUp, \#Educateb4uVax, \#forcedvaccines, NoForcedVaccines, \#NoFORCEDvaccines, \#NoVaccineMandates, notomandatoryvaccines, NoVaccine, \#NoVaccineForMe, \#DoNotVaccinateTheChildren, \#VaccineFailure, \#VaccineSideEffects, novaccinemandates, saynotovaccines, stopmandatoryvaccination, unvaccinated, vaccinationchoice, VaccineAgenda, vaccinedamage, vaccinefailure, vaccinefraud, vaccineharm, vaccineinjuries, vaccineinjury, VaccinesAreNotTheAnswer, vaccinesarepoison, vaccinescause, vaccineskill, vaxxed, vaccine, vaccination, vaccinate, vaccinating, vaccinated, antivaxer, antivaxers, antivaxxer, \#Igetvaccinated, \#COVIDvaccine, \#COVID19Vaccine, \#covax, \#VaccinesSaveLives, \#pfizer, \#ReadyToVaccinate, \#CDCwhistleblower, \#VaccinesWork, \#CoronaVirusVaccine, \#Vaccinated, \#unsafevaccines, \#antivaxx, \#NoVaccine, \#StopVaccine, \#AntiVaccine, \#AntiVacc, \#Covaxin, \#Covishield, \#GotTheDose, \#VaccinationDrive, \#VaccineForIndia, \#We4Vaccine, \#LargestVaccinationDrive, \#pfizer, \#VaccineSideEffects, \#VaccineDeath, \#VaccineInjuries, \#vaccinegenocide, \#vaccinedeaths, vaccines, vaccinations, \#vaccine, \#vaccines, \#vaccination, \#vaccinations, \#coronavaccine, \#CovidVaccineScam, \#covidvaccinedeaths, \#covidvaccineVictims, \#CovidVaccineKills, \#BanCovidVaccine, \#vaxxdamage, \#CovidVaccinesKill, notomandatory vaccines, Covishield- side effects, covishield-side effects, SputnikV unsafe, abolishbigpharma, notocoronavirusvaccines, getvaccinated, iwillgetvaccinated, Covaxin affect, COVISHIELD affect, pfizervaccine affect, \#pfizerkills, \#modernavaccine, \#SputnikVaccine, \#Oxfordvaccine, \#astrazenica, \#VaccineSaveLives, \#VACCINESWORK, covid19vaccine, covid19 pfizer, covid19 moderna, covid19 astrazeneca, covid19 biontech, pfizercovidvaccine, covidvaccine pfizer, modernacovidvaccine, astrazenecacovidvaccine, biontechcovidvaccine, covidvaccine, coronavirusvaccine, coronavaccine, vaccinessavelives, pfizervaccine, modernavaccine, \#vaccinemandates, \#Sinovac, \#BharatBiotech, \#COVIDBooster, \#COVIDVaccineBooster, \#boosterdose, \#VaccineBooster, vaccine booster shot, \#nasalvaccine \\ \hline
\end{tabular}
    \label{keywords-table}
\end{table}

\subsubsection{The \textit{PAIHcovax} Dataset}
\textit{GeoCovaxTweets Extended} contains 2.8 million tweets, so manual annotation is not feasible. We randomly select a smaller subset of tweets to address this issue for manual annotation. The randomly sampled tweets were then annotated, and we refer to that subset of tweets as the \textit{PAIHcovax} dataset. The dataset has 4,950 labelled English-language tweets. This dataset is larger than other vaccines and vaccination-related English-only tweets datasets (as shown in Table \ref{multidata}). We manually annotate 4,950 tweets into five classes, namely ``Pro-covidvax'', ``Anti-covidvax'', ``Informative-covidvax'', ``Hesitant-covidvax'' and ``Not-related''. We follow the annotation approach adopted by many prior studies \citep{chen2022multilingual,mu2023vaxxhesitancy,poddar2022winds}. Additionally, we gather common myths and facts surrounding COVID-19 vaccines and vaccination from reputable sources such as Healthline\footnote{\url{https://www.healthline.com/health-news/doctors-debunk-9-popular-covid-19-vaccine-myths-and-conspiracy-theories}}, Healthcare \footnote{\url{https://www.muhealth.org/our-stories/covid-19-vaccine-myths-vs-facts}}, Public Health\footnote{\url{https://www.publichealth.org/public-awareness/understanding-vaccines/vaccine-myths-debunked/}}, the Centers for Disease Control and Prevention (CDC)\footnote{\url{https://www.cdc.gov/coronavirus/2019-ncov/vaccines/facts.html}}, and Health\footnote{\url{https://www.hopkinsmedicine.org/health/conditions-and-diseases/coronavirus/covid-19-vaccines-myth-versus-fact}}. These sources are used to curate a list of ground truths for tagging tweets as ``Pro-covidvax'' and ``Anti-covidvax''. This compilation encompass myths such as `the vaccine can alter DNA', `the vaccine can cause infertility', `the vaccine contains dangerous toxins', and `the vaccine contains tracking devices', among others. For the ``Hesitant-covidvax'' class, we adopt the concept of ``vaccine hesitancy'', which refers to the resistance towards accepting vaccines even when vaccination services are readily accessible and available \cite{macdonald2015vaccine}. Furthermore, we identify hesitant tweets as those discussing minimal risks associated with vaccine-preventable diseases, leading to the belief that vaccination is not essential\footnote{\url{https://www.hopkinsmedicine.org/health/conditions-and-diseases/coronavirus/covid19-vaccine-hesitancy-12-things-you-need-to-know}}. Tweets are labelled as ``Informative-covidvax'' if they contain information about vaccine-related news, conditions for vaccination, government vaccination plans, vaccination progress and reports, as well as advice or instructions from official Twitter accounts of health agencies and governmental organizations. 

We employ three independent annotators with Master's degree qualifications and proficiency in the English language for the annotation process. The tweets are distributed among these three independent annotators, ensuring that each tweet is annotated by at least two annotators. We use Label Studio\footnote{\url{https://labelstud.io/}}, an open-source labelling tool, for annotation because of its user-friendly and easy-to-learn interface. To aid in understanding the class labels, we provide sample tweets with gold labels to the annotators. After annotation, the dataset had 2393 ``Pro-covidvax'', 1654 ``Anti-covidvax'', 664 ``Informative-covidva'', 87 ``Hesitant-covidvax'', and 152 ``Not-related'' tweets. We compute Cohen’s kappa coefficient to measure inter-annotator agreement among the independent annotators, and we report the statistic to be 0.78, which indicates an almost-perfect agreement. Lastly, we invited medical domain experts in public health to validate the annotation process. This approach ensured the accuracy of manual data annotation and maintained a high standard of dataset quality. Example tweets across the 5 classes are provided in Table \ref{sample-tweets}.
\begin{table}
\scriptsize
    \centering
    \caption{A sample of tweets from \textit{PAIHcovax}.}
    \label{sample-tweets}
    \begin{tabular}{c|l}
\hline
\textbf{Class} &
  \multicolumn{1}{c}{\textbf{Examples}} \\ \hline
\textbf{\begin{tabular}[c]{@{}c@{}}Pro-\\covidvax\end{tabular}} &
  \begin{tabular}[c]{@{}l@{}}\textbf{E1:} The low risk of transmission outdoors, combined with the very effective\\COVID vaccines, means vaccinated Americans no longer need to wear\\masks when outside, even when meeting unvaccinated friends and family.\\ \\  \textbf{E2:} You can safely end masking requirements today for anyone who is two \\weeks post second vaccine dose. this should be done, because it will encourage\\unvaccinated to get vaccines, and it will start to normalize the return to non-mask\\wearing.\end{tabular} \\ \hline
\textbf{\begin{tabular}[c]{@{}c@{}}Anti-\\covidvax\end{tabular}} &
\begin{tabular}[c]{@{}l@{}}\textbf{E1:} No, It's not safe. Do NOT go Seriously. I'm in a similar situation \& my\\doc has told me that no vaccine is 100\% effective against all variants.\\ The 1 I got doesn't do well against the worst variants. \\ \\ \textbf{E2:} Vaccinations are making Covid worse: "The resurgence in both hospitalisations\\and deaths is dominated by those that have received two doses of the vaccine...\\ immunization failures account for more serious illness than unvaccinated individuals."\\\end{tabular} \\ \hline

\textbf{\begin{tabular}[c]{@{}c@{}}Infomative-\\covidvax\end{tabular}} &
  \begin{tabular}[c]{@{}l@{}}\textbf{E1:} What does it mean a vaccine is 95\% effective? A vaccinated population\\will have 95\% less cases than unvaccinated. U.S. cases = 30 Mil, Approx\\1 of every 11 infected, Plus approx 25\% of the infected went undiagnosed. So approx\\1 of 9 infected. So 95\% protection = approx 1 out of 95\\ \\ \textbf{E2:} Real-world data shows the efficacy of Pfizer/BioNTech vaccine. Pfizer/BioNTech\\COVID-19 vaccine was effective in a study conducted in Israel involving nearly\\600,000 pairs of matched vaccinated and unvaccinated people.\\ The vaccine was 92\% effective in preventing severe disease \end{tabular} \\ \hline
\textbf{\begin{tabular}[c]{@{}c@{}}Hesitant-\\covidvax\end{tabular}} &
  \begin{tabular}[c]{@{}l@{}}\textbf{E1:} We still don't know how long the vaccine will last? Will you need a booster\\ in 2 yr? Dont know. WE still don't know if the vaccine is effective in preventing\\ a vaccinated person from spreading covid to a non vaccinated person. Im vaccinated.\\ Can I hug my unvaccinated mom? Don’t know.\\ \\ \textbf{E2:} will the vaccine make me feel like I am now in a safer portal\\and away from these unvaccinated people?\end{tabular} \\ \hline
\textbf{\begin{tabular}[c]{@{}c@{}}Not-\\related\end{tabular}} &
  \begin{tabular}[c]{@{}l@{}} \textbf{E1:} My colleagues daughter received 1st dose of HPV vaccine\\in 2019 \& never received an update on the 2nd. Will it be effective if it is\\given 2 yrs on? Will they even get it or are they just classed as\\unvaccinated now as it wasn't completed?\\ \\ \textbf{E2:} A new study has shown the effectiveness of the HPV vaccine, and\\found a dramatic decline in human papillomavirus infections in both\\vaccinated and unvaccinated teen girls and young women in the United States.\\ HTTPURL
  \end{tabular} \\ \hline
\end{tabular}
\end{table}

\subsubsection{Upsampling}

\begin{figure}
     \centering
     \begin{subfigure}[b]{0.49\textwidth}
         \centering
         \includegraphics[width=1\textwidth]{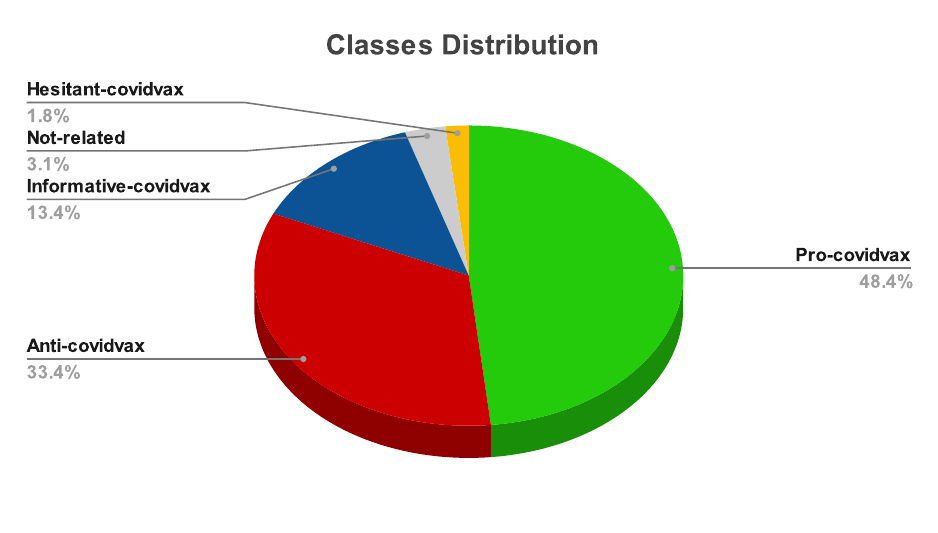}
        \caption{}
        \label{classes}
     \end{subfigure}
     \hfill
     \begin{subfigure}[b]{0.49\textwidth}
         \centering
         \includegraphics[width=1\textwidth]{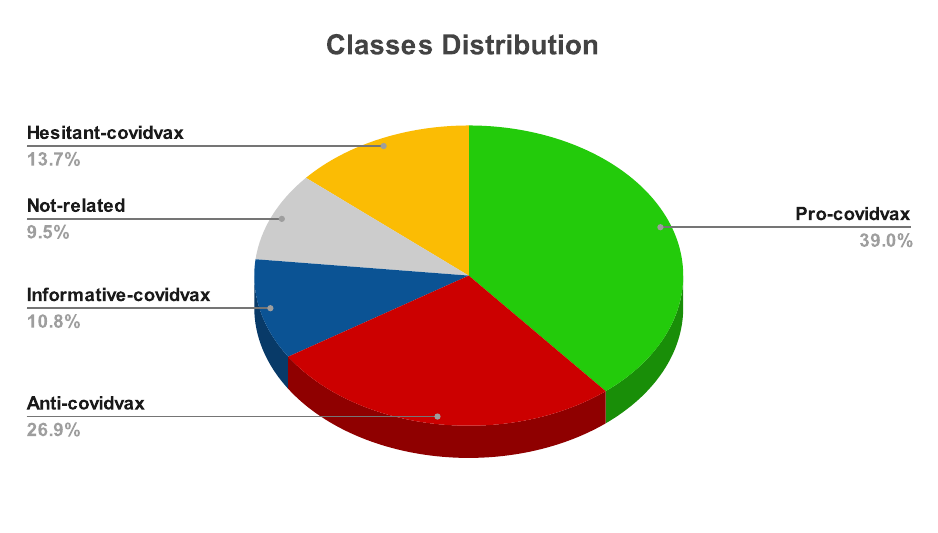}
        \caption{}
        \label{upsampling}
     \end{subfigure}
\caption{Class-wise distribution of tweets in \textit{PAIHcovax} before (Sub-figure a) and after (Sub-figure b) upsampling.}
\end{figure}

In \textit{PAIHcovax}, hesitant and not-related classes contain a relatively small sample of tweets compared to the other three classes, as indicated in Figure \ref{classes}. This could present a challenge when fine-tuning language models, as they might exhibit bias towards the majority class and have difficulty properly categorizing samples from the minority class. One way to address this imbalance is to reduce the number of examples in the majority class by undersampling it. However, this method could lose important information and hinder the model's ability to learn from the available data. Instead, oversampling the minority class by generating additional examples could better balance the dataset. However, duplicating existing samples might lead to overfitting and limit the model's generalization capability. Therefore, data augmentation techniques such as synonym replacement, paraphrasing, and back-translation can generate additional samples for the minority classes and help the model recognize them better. Our study used a Text-to-Text Transfer Transformer (T5) model \citep{raffel2020exploring} to generate new paraphrased tweets from the original tweets in hesitant and not-related classes of \textit{PAIHcovax}. Figure \ref{upsampling} shows the class distribution after upsampling.

\subsection{CovaxBERT}
Pre-trained language models and libraries are widely used for short text classification, including stance classification. These models can be fine-tuned on specific tasks with minimal data, thanks to transfer learning, which has revolutionized NLP. Transformers, a neural network architecture introduced in \citep{vaswani2017attention}, process sequential data, such as text, and are highly effective for various NLP tasks. The most popular transformer-based model is BERT \citep{devlin-etal-2019-bert}, which uses a pre-training method called masked language modeling to capture contextual information. Other transformer-based models, such as RoBERTa \citep{liu2019roberta}, XLNet \citep{yang2019xlnet}, ELECTRA \citep{clark2020electra}, and BERTweet \citep{nguyen-etal-2020-bertweet}, have been developed to achieve even better performance on various NLP tasks. Overall, transformers and their variations have become essential for NLP research, leading to significant advancements in the field.

In this study, we fine-tune multiple pre-trained transformer-based models, namely BERT \citep{devlin-etal-2019-bert}, DistilBERT \citep{sanh2019distilbert}, RoBERTa \citep{liu2019roberta}, XLM-RoBERTa \citep{conneau2019unsupervised}, ELECTRA \citep{clark2020electra}, BERTweet \citep{nguyen-etal-2020-bertweet}, and XLNet \citep{yang2019xlnet} on the annotated dataset (\textit{PAIHcovax}) using the Hugging Face library \citep{wolf-etal-2020-transformers} as \textit{CovaxBERT} candidates. The best-performing model in terms of F1-score is selected as \textit{CovaxBERT} for analyzing \textit{GeoCovaxTweets Extended}. Our approach aligns with the methodology presented in \cite{nguyen2020bertweet}, where a classification head is introduced to the pre-trained models for the purpose of fine-tuning, and the F1 score (macro) is reported.

\section{Dataset Description}
\label{exploring-dataset}
Geotagged data plays an important role in modelling location-specific information \citep{comito2021covid}. In particular, spatial analyses benefit from geographical data based on the origin locations of tweets rather than the toponyms in tweet texts. This is because dependence on toponyms can generate biased results due to the Location A/B problem, where people in one location may participate in a discourse specific to another location \citep{lamsal2022addressing,10020460}. Consequently, two types of geographical metadata are available for tweets: ``tweet location," where a user shares their location when creating a tweet, either as exact geocoordinates or a bounding box, and ``account location," which is based on the user's public profile location. The \textit{GeoCovaxTweets Extended} Dataset contains only the tweets with point geocoordinates or bounding boxes because Twitter does not validate the account location field. Today, $<$1\% of tweets are geotagged; however, a comparative analysis by \citep{lamsal2022twitter} reported the daily distributions of full-volume (based on Twitter’s counts API) and geotagged tweets to have significantly identical patterns. Therefore, the volumetric analysis performed on \textit{GeoCovaxTweets Extended} can be considered a near-true interpretation of the full-volume tweets.

\begin{figure}
     \centering
     \begin{subfigure}[b]{0.49\textwidth}
         \centering
         \includegraphics[width=1\textwidth]{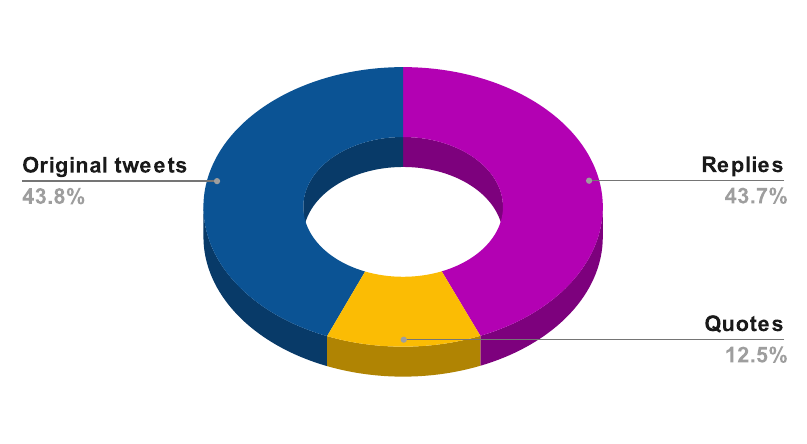}
        \caption{Distribution of original, reply, and quote tweets in \textit{GeoCovaxTweets Extended}.}
        \label{referenced}
     \end{subfigure}
     \hfill
     \begin{subfigure}[b]{0.49\textwidth}
         \centering
         \includegraphics[width=1\textwidth]{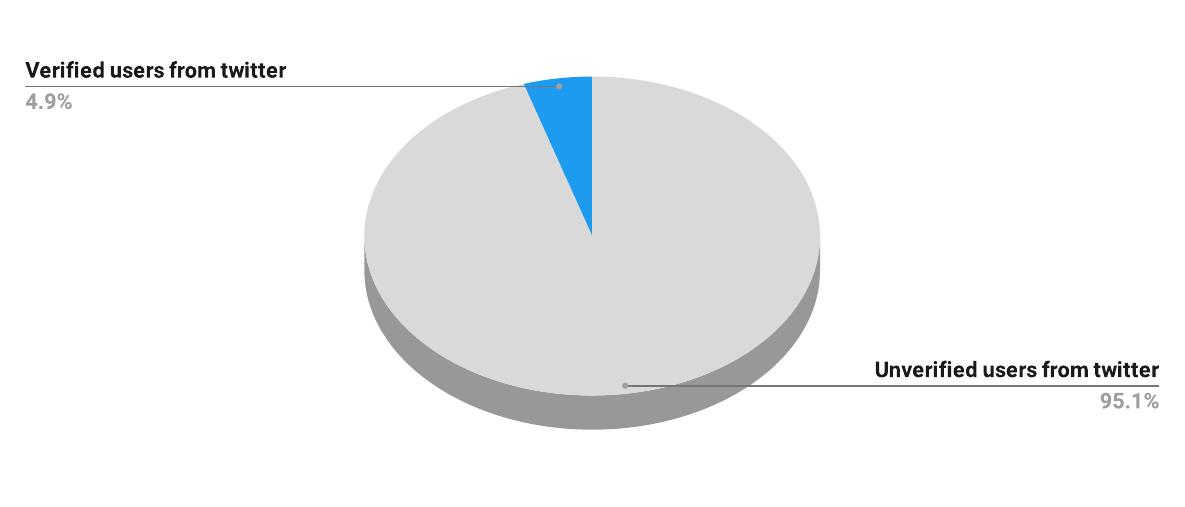}
         \caption{Distribution of tweets from verified versus not verified users in \textit{GeoCovaxTweets Extended}.}
         \label{verifiedUsers}
     \end{subfigure}
\caption{}
\end{figure}

\textit{GeoCovaxTweets Extended} comprises 2.8 million tweets regarding COVID-19 vaccines and vaccination, generated by 650,173 unique users from 238 countries worldwide between January 2020 and March 2023. Various aspects of the dataset are analyzed in this section, including overview (in Table \ref{statistic}), country distribution (in Table \ref{top countries}),languages (in Table \ref{multidata}). Statistics reveal that most tweets in the discourse are from unverified users (as shown in Figure \ref{verifiedUsers}). Figure \ref{referenced} shows the proportion of original, reply, and quote tweets. Specifically, 43.8\% of the tweets are original, 43.7\% are reply tweets, and 12.5\% are quote tweets.

\begin{table}
\caption{Overview of \textit{GeoCovaxTweets Extended}.}
\label{statistic}
\footnotesize
\centering
    \begin{tabular}{c|c}
    \hline
    Total Tweets & 2,812,332 \\
    Unique Tweets & 2,797,258 \\ 
    Users & 650,173 \\
    URLs (unique) & 1,088,733 \\
    Mentions (unique) & 810,144 \\
    Hashtags (unique) & 178,918 \\
   Languages & 62 and undefined \\ 
    Countries \& Territories & 238 \\
    \hline
   \end{tabular}
\end{table}

\begin{table}
\caption{Most frequent languages (N = 62, $^a$ISO 639-1 two-letter language codes) in \textit{GeoCovaxTweets Extended}.}
\label{multidata}
\footnotesize
\centering
\begin{tabular}{ccc}
\hline
\multicolumn{1}{c|}{\textbf{Language}}                                                                                                          & \multicolumn{1}{c|}{\textbf{ISO$^a$}}                                                                     & \textbf{\# of Tweets}                                                                    \\ \hline
\multicolumn{1}{c|}{English}                                                                                                                    & \multicolumn{1}{c|}{en}                                                                               & 2,491,947                                                                                  \\ 
\multicolumn{1}{c|}{French}                                                                                                                     & \multicolumn{1}{c|}{fr}                                                                               & 85,147                                                                                    \\ 
\multicolumn{1}{c|}{Hindi}                                                                                                                      & \multicolumn{1}{c|}{hi}                                                                               & 29,442                                                                                    \\ 
\multicolumn{1}{c|}{Tagalog}                                                                  & \multicolumn{1}{c|}{tl}                                                                               & 22,810                                                                                    \\ 
\multicolumn{1}{c|}{Spanish}                                                                                                                    & \multicolumn{1}{c|}{es}                                                                               & 20,590                                                                                    \\ 
\multicolumn{1}{c|}{Indonesian}                                                                                                                 & \multicolumn{1}{c|}{in}                                                                               & 18,433                                                                                    \\ \hline
\multicolumn{3}{c}{\begin{tabular}[c]{@{}c@{}}other languages in order to their frequency:\\ it, pt, ja, nl, de, da, et, th, ca, tr, ar, mr, gu, ro,\\ ht, sv, zh, ta, pl, fi, bn, el, ur, te, vi, cs, or, ne,\\ ru, kn, iw, sl, ko, lv, no, ml, lt, si, fa, hu, cy, \\ is, eu, pa, uk, sr, bg, km, dv, ps, sd, ka, lo, hy,\\ am, bo\end{tabular}} \\ \hline
\end{tabular}

\end{table}

\begin{table}
\caption{List of the top 50 countries in \textit{GeoCovaxTweets Extended}.}
\label{top countries}
    \centering
    \footnotesize
\begin{tabular}{c|c|c|c|c|c}
\hline
\textbf{\textbf{SN}} & \textbf{\textbf{Country}} & \textbf{\textbf{Frequency}} & \textbf{\textbf{SN}} & \textbf{\textbf{Country}} & \textbf{Frequency} \\ \hline
1 & United States & 1,326,222 & 26 & Uganda & 6,347 \\ \hline
2 & United Kingdom & 342,508 & 27 & Indonesia & 5,836 \\ \hline
3 & India & 253,478 & 28 & Switzerland & 5,399 \\ \hline
4 & Canada & 204,446 & 29 & Saudi Arabia & 5,352 \\ \hline
5 & Australia & 97,926 & 30 & Ghana & 5,186 \\ \hline
6 & France & 68,977 & 31 & Denmark & 4,692 \\ \hline
7 & South Africa & 58,691 & 32 & Sweden & 4,604 \\ \hline
8 & Ireland & 42,620 & 33 & Singapore & 4,403 \\ \hline
9 & Philippines & 40,303 & 34 & Israel & 4,341 \\ \hline
10 & Malaysia & 31,829 & 35 & Trinidad and Tobago & 3,756 \\ \hline
11 & New Zealand & 18,117 & 36 & Turkey & 3,564 \\ \hline
12 & Spain & 15,762 & 37 & Botswana & 3,423 \\ \hline
13 & Netherlands & 15,296 & 38 & Argentina & 3,417 \\ \hline
14 & Pakistan & 14,694 & 39 & Sri Lanka & 3,203 \\ \hline
15 & Nigeria & 13,955 & 40 & Colombia & 3,017 \\ \hline
16 & Germany & 13,072 & 41 & Zimbabwe & 2,954 \\ \hline
17 & Italy & 11,503 & 42 & China & 2,818 \\ \hline
18 & Kenya & 10,793 & 43 & Greece & 2,576 \\ \hline
19 & Thailand & 10,102 & 44 & Portugal & 2,544 \\ \hline
20 & Brazil & 9,341 & 45 & Nepal & 2,270 \\ \hline
21 & Japan & 8,758 & 46 & Poland & 2,255 \\ \hline
22 & Mexico & 8,608 & 47 & Maldives & 2,167 \\ \hline
23 & Belgium & 8,525 & 48 & Chile & 2,141 \\ \hline
24 & Jamaica & 7,326 & 49 & Bangladesh & 2,052 \\ \hline
25 & United Arab Emirates & 6,477 & 50 & Lebanon & 2,035 \\ \hline
\end{tabular}
\end{table}

\begin{figure}
\centering
\includegraphics[width=1.0\textwidth]{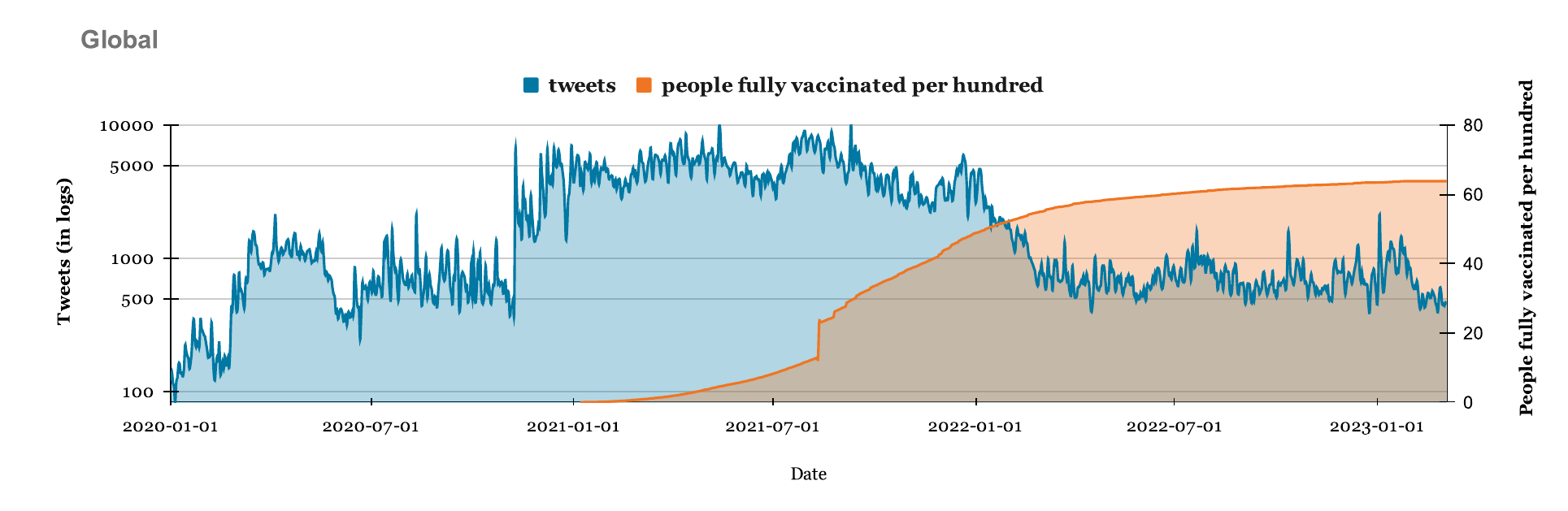}
\caption{ The daily distribution of tweets and the cumulative number of people fully vaccinated per hundred (globally).}
\label{daily-dist}
\end{figure}
\begin{figure}
     \centering
     \begin{subfigure}[b]{0.49\textwidth}
         \centering
         \includegraphics[width=0.86\textwidth]{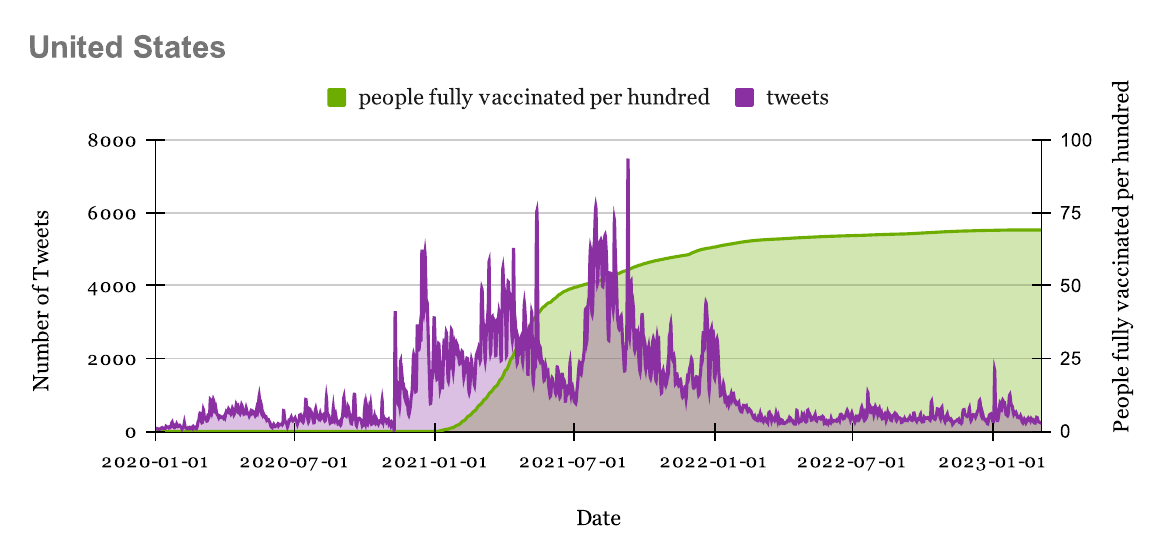}
         \label{fig:1}
     \end{subfigure}
     \hfill
     \begin{subfigure}[b]{0.49\textwidth}
         \centering
         \includegraphics[width=0.86\textwidth]{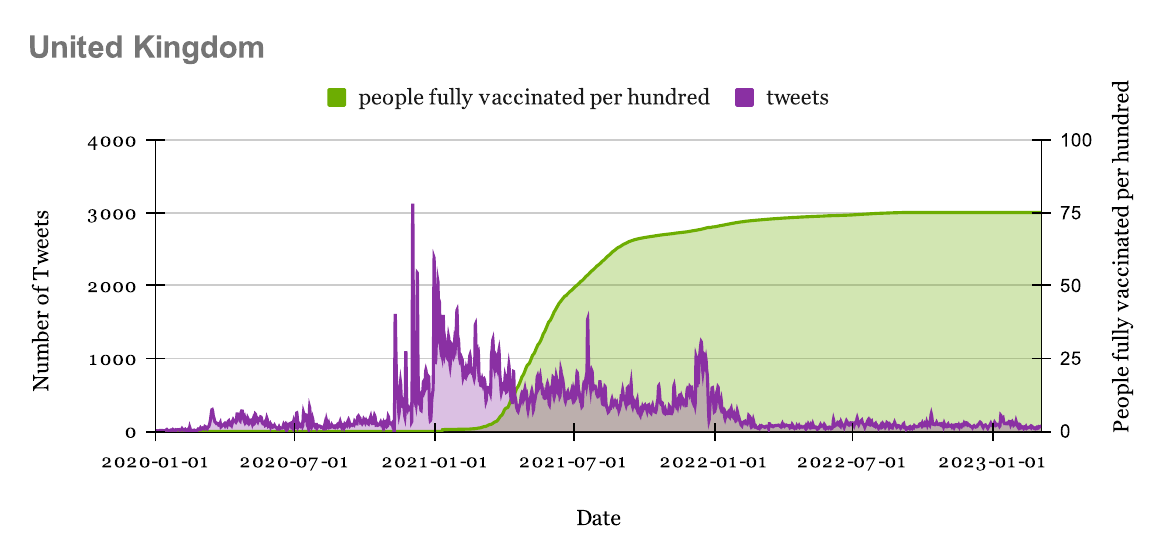}
         \label{fig:2}
     \end{subfigure}\\
     
     \begin{subfigure}[b]{0.49\textwidth}
         \centering
         \includegraphics[width=0.86\textwidth]{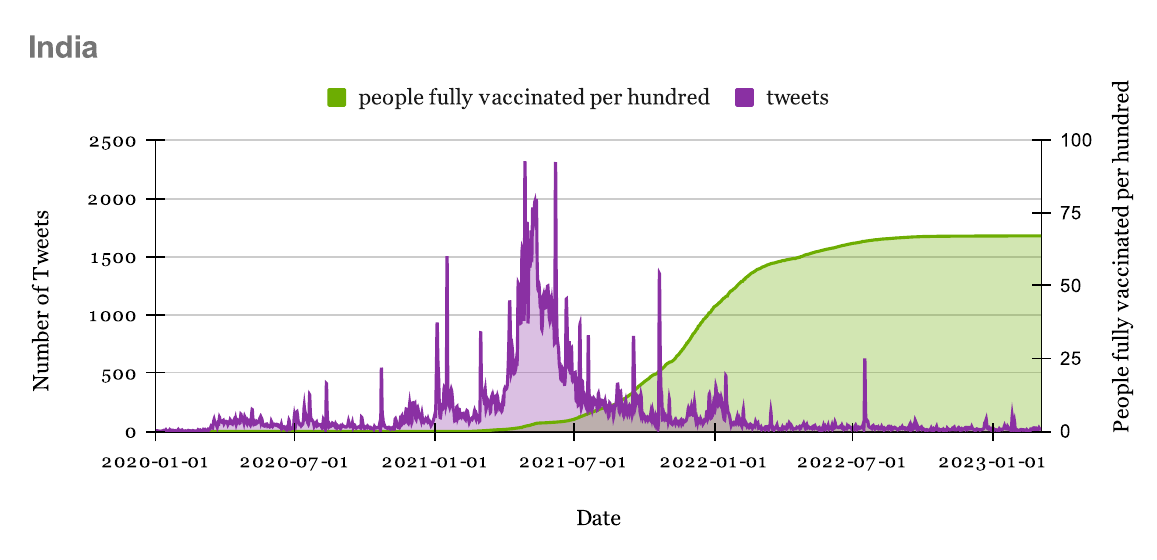}
         \label{fig:3}
     \end{subfigure}
     \hfill
     \begin{subfigure}[b]{0.49\textwidth}
         \centering
         \includegraphics[width=0.86\textwidth]{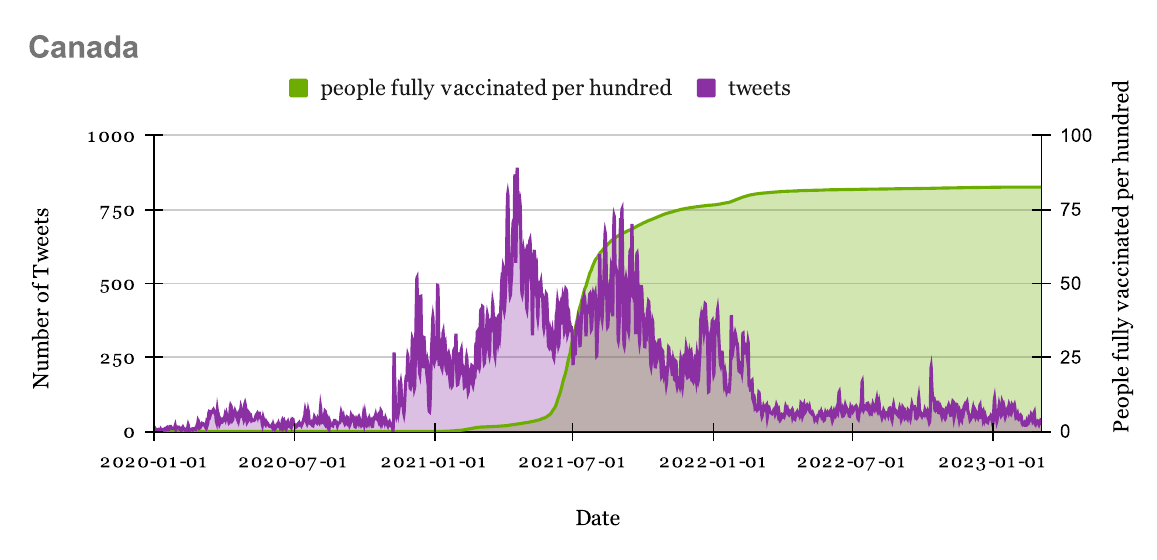}
         \label{fig:4}
     \end{subfigure}\\
     
     \begin{subfigure}[b]{0.49\textwidth}
         \centering
         \includegraphics[width=0.86\textwidth]{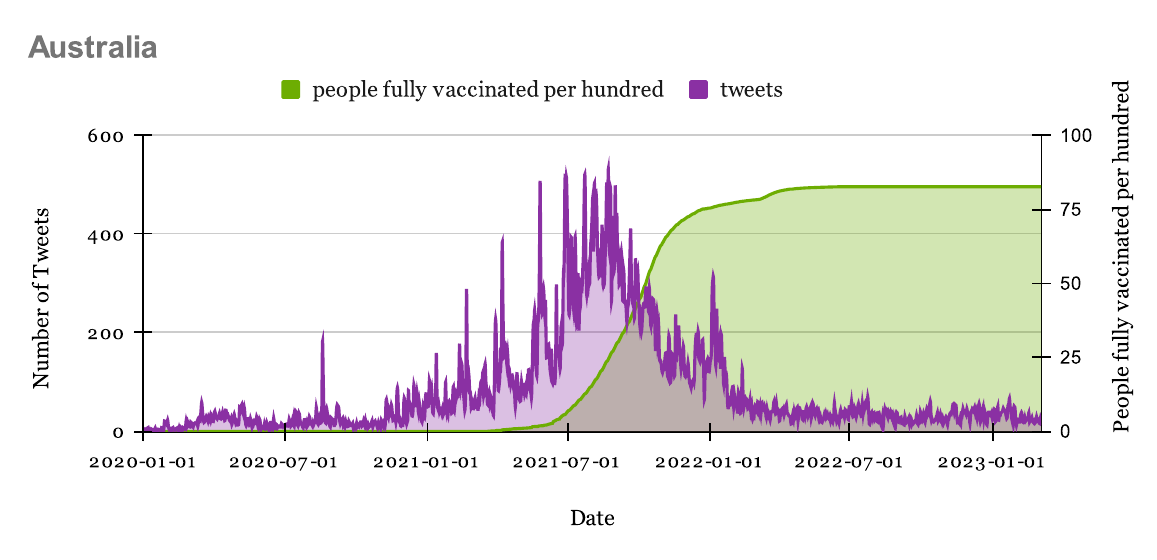}
         \label{fig:5}
     \end{subfigure}
     \hfill
     \begin{subfigure}[b]{0.49\textwidth}
         \centering
         \includegraphics[width=0.86\textwidth]{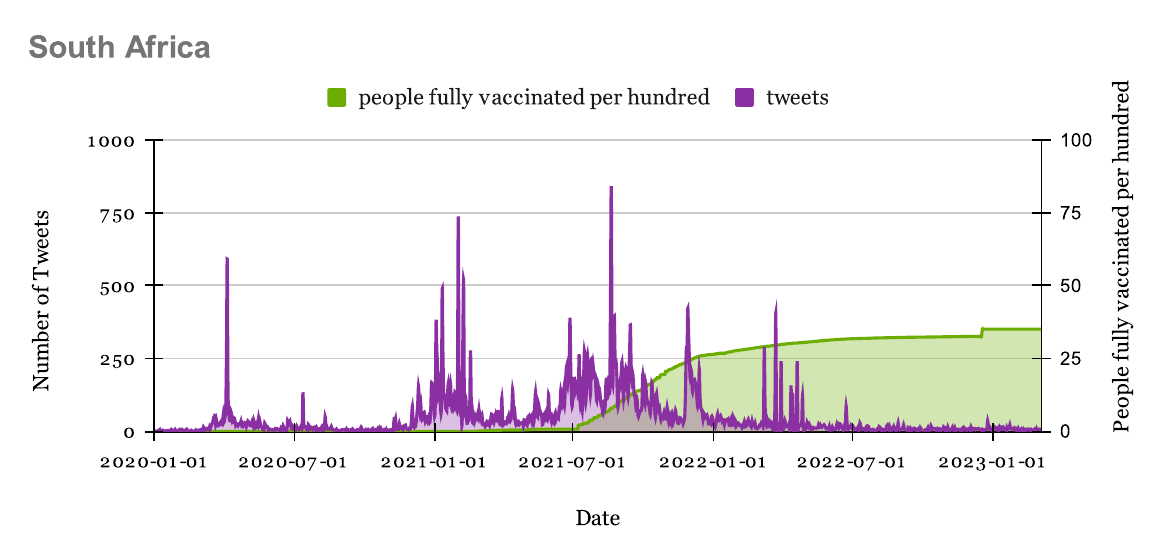}
         \label{fig:6}
     \end{subfigure}
        
\caption{The daily distribution of tweets in the top six countries involved in the discourse and their cumulative number of people fully vaccinated per hundred.}
\label{six-graphs}
\end{figure}

The daily distribution of tweets globally and the cumulative number of fully vaccinated people per hundred, illustrated in Figure \ref{daily-dist}, offer interesting insights into the evolution of attitudes towards discourse worldwide. The Pfizer-BioNTech vaccine was the first COVID-19 vaccine to receive approval for emergency use, as stated in a report by the WHO\footnote{https://www.who.int/news/item/31-12-2020-who-issues-its-first-emergency-use-validation-for-a-covid-19-vaccine-and-emphasizes-need-for-equitable-global-access} on December 2, 2020. Following its approval, social media activity concerning people's opinions and discussions about the vaccine increased significantly. This implies that people were more interested in discussing and sharing their opinions on the vaccine, whether positive, negative, or hesitant.

Social media discourse on COVID-19 vaccines and vaccination vary significantly from country to country, depending on cultural attitudes towards vaccination, government policies and regulations, and public health infrastructure. The top six countries contributing the highest number of English-language tweets in \textit{GeoCovaxTweets Extended} are the United States, the United Kingdom, India, Canada, Australia, and South Africa. Figure \ref{six-graphs} displays the daily distribution of tweets in these six countries and their cumulative number of people fully vaccinated per hundred. The source for vaccination data in this study is \textit{Our World in Data} \citep{mathieu2021global}.

\section{Findings and Discussions}

The rollout of COVID-19 vaccines began in late 2020 and continued into 2021, with the initial focus being on vaccinating high-risk groups and frontline workers. According to \textit{Our World in Data}\footnote{https://ourworldindata.org/covid-vaccinations}, as of March 2023, over 5.55 billion people have been fully vaccinated. However, vaccination rates have been uneven across different regions and countries, with some countries lagging behind due to supply constraints, vaccine hesitancy, or other factors. Moreover, new challenges emerged in 2022, such as the emergence of new variants, prompting some countries to provide booster shots to their populations, adding another layer of complexity to the vaccination effort as countries work to balance the need to provide additional protection against the virus ensuring vaccines are available to those who have not yet been vaccinated. Hence, to analyze the behaviour of Twitter users towards COVID-19 vaccines and vaccination, we used \textit{CovaxBERT} to analyze English-language tweets from \textit{GeoCovaxTweets Extended}. We focus our analysis on specific tweet objects: date/time, tweet text, tweet source, followers, verification status, tweet count, country, and hashtags.

\begin{figure}
\centering
\includegraphics[width=0.8\textwidth]{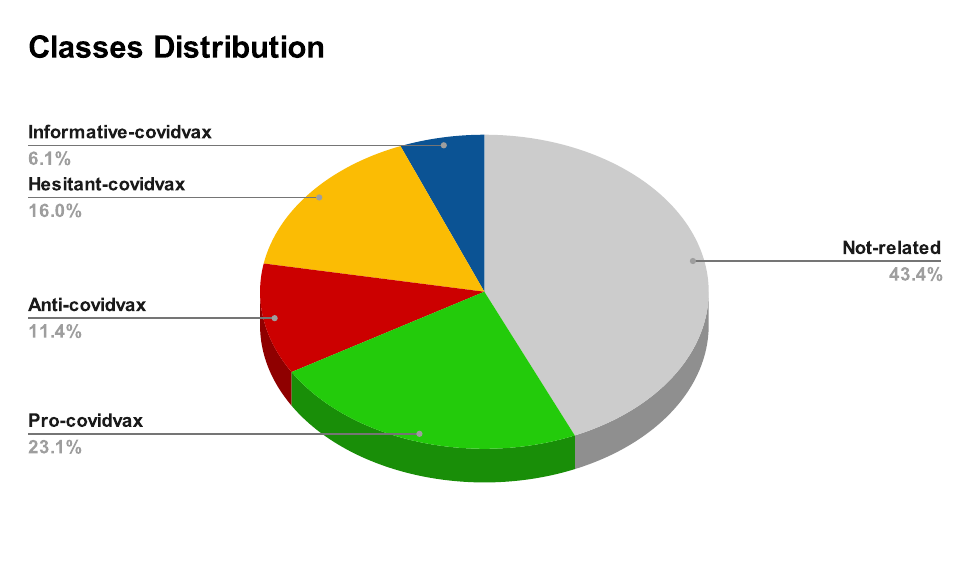}
\caption{The class-wise distribution of the 2.4m English language tweets in \textit{GeoCovaxTweets Extended}.}
\label{daily-dist per category}
\end{figure}

We input every English-language tweet in \textit{GeoCovaxTweets Extended} to \textit{CovaxBERT}. The results are illustrated in Figure \ref{daily-dist per category}, indicating that 43.4\% of the tweets were irrelevant to the discourse on COVID-19 vaccines or vaccination. Such tweets lacked significant information and addressed other vaccine types or pandemic aspects. Examples of such tweets are provided in Appendix A. Further, the distribution was as follows: Pro-covidvax (23.1\%), Hesitant-covidvax (16.0\%), Anti-covidvax (11.4\%), and Informative-covidvax (6.1\%). We considered the four class of tweets (Pro-covidvax, Hesitant-covidvax, Anti-covidvax, and Informative-covidvax) for further analysis related to the discourse. Several tweet distributions are drawn based on these classes and discussed next.

\begin{table}
\tiny
\caption{Top 20 hashtags across each class.}
\centering
\label{classwisetop20hashtags}
\scriptsize
\begin{tabular}{p{3cm}|p{10cm}}
\hline
\textbf{Class} & \textbf{Hashtags} \\
  \hline
\multirow{6}{*}{\textbf{Pro-covidvax}} & \#covid19 (22,324), \#covidvaccine (8,986),   \#getvaccinated (8,187), \#vaccine (8,122), \#vaccinated (6,914), \#covid (5,777),   \#vaccination (3,315), \#vaccineswork (2,881), \#covid19vaccine (2,501), \#vaccines   (2,329), \#covid\_19 (2,178), \#pfizer (2,151), \#covidvaccination (1,837),   \#coronavirus (1,806), \#coronavirus (1,806), \#coronavirus (1,806), \#wearamask   (1,790), \#getvaccinatednow (1,670), \#vaccinessavelives (1,537), \#maskup (1,363),   \#nhs (1,095), \#astrazeneca (1,049)\\
\hline
\multirow{5}{*}{\textbf{Anti-covidvax}} & \#covid19 (4,522), \#vaccine (2,294), \#covid   (1,672), \#antivaxxers (1,657), \#covidvaccine (1,384), \#vaccines (1,202),   \#getvaccinated (881), \#vaccinated (680), \#coronavirus (599), \#pfizer (590),   \#bigpharma (585), \#covid\_19 (575), \#covidiots (584), \#vaccination (511),   \#trump (404), \#vaccinesideeffects (352), \#vaccineswork (351), \#unvaccinated   (312), \#cdnpoli (275), \#plandemic (261) \\
\hline
\multirow{5}{*}{\textbf{Informative-covidvax}} & \#covid19 (14,575), \#vaccine (4,249), \#covid   (2,421), \#coronavirus (2,410), \#vaccination (1,866), \#pfizer (1,427),   \#covid19vaccine (1,416), \#vaccines (1,291), \#getvaccinated (1,246), \#covid\_19   (1,165), \#vaccinated (861), \#breaking (799), \#astrazeneca (692), \#india (675),   \#covaxin (658), \#indiafightscorona (585), \#wearamask (574), \#covishield   (546), \#vaccineswork (532) \\
 \hline
\multirow{5}{*}{\textbf{Hesitant-covidvax}} & \#covid19 (8,636), \#vaccine (5,458),   \#covidvaccine (4,829), \#vaccinated (3,655), \#getvaccinated (3,434), \#vaccination   (2,488), \#covid (2,029), \#vaccineswork (1832), \#pfizer (1,794), \#covid\_19   (1,489), \#vaccines (1,348), \#covid19vaccine (1,196), \#coronavirus (1,087),   \#wearamask (772), \#coronavaccine (770), \#vaccinessavelives (746),   \#antivaxxers (739), \#moderna (732), \#astrazeneca (603), \#bigpharma (576) \\
 \hline
\end{tabular}
\end{table}

Table \ref{classwisetop20hashtags} presents information about the top 20 hashtags and their frequency across each class. Notably, the hashtag \#covid19 is the most common across all classes, indicating that the discourse concerns COVID-19 and its vaccines and vaccination context. However, some hashtags are present in multiple classes, with differences in their frequency suggesting that the way people participate in the discourse varies depending on their stance. We observed the presence of contextual hashtags in Pro-covidvax, and Anti-covidvax classes only. In the Pro-covidvax class, we observe the presence of contexual hashtags such as \#getvaccinated, \#vaccineswork, \#vaccinessavelives, etc., that indicate support for measures to combat COVID-19 through vaccinations. In the Anti-covidvax class, there is the presence of contextual hashtags such as \#vaccinesideeffects and \#plandemic that negatively correlate with the vaccination context. In the Informative-covidvax class, most tweets seem to be related to news or information about COVID-19 vaccines. This class' only top contextual hashtag seems to be \#getvaccinated, \#breaking, \#indiafightscorona, and \#vaccineswork. The hesitant class, however, does not seem to contain contextual hashtags.

Based on the tweet distribution across countries and each class (shown in Table \ref{classwisetopcountries}), it is evident that native English-speaking nations dominate as only English tweets are analyzed. The tweet distribution across countries ranks the United States with the most tweets across all classes. The United States generated 304k Pro-covidvax, 209k Hesitant-covidvax, 174k Anti-covidvax, and 64k Informative-covidvax tweets. The United Kingdom is ranked second with 74k Pro-covidvax, 54k Anti-covidvax, 38k Hesitant-covidvax tweets, and 17k Informative-covidvax tweets. Similarly, India ranks fourth in terms of Pro-covidvax and Hesitant-covidvax, fifth in terms of Anti-covidvax, and second in terms of Informative-covidvax. A comprehensive distribution of the top 20 countries across each class in provided in Table \ref{classwisetopcountries}.

\begin{table}
\centering
\caption{Top 20 Countries along with their frequency across each class.}
\scriptsize
\label{classwisetopcountries}
\begin{tabular}{p{3.5cm}|p{9cm}}
\hline
\textbf{Class}      & \textbf{Countries} \\ \hline
\multirow{5}{*}{\textbf{   Pro-covidvax }}       & United States (304,532), United Kingdom   (74,906), Canada (48,006), India (44,157), Australia (21,489), Ireland (9,969),   South Africa (9,459), Republic of the Philippines (5,353), New Zealand (4,326),   Malaysia (3,579), Pakistan (3,332), Nigeria (2,435), Germany (2,297), Kenya   (1,979), Jamaica (1,656), Spain (1,535), Thailand (1,526), France (1,493), Uganda   (1,312), United Arab Emirates (1,262)                          \\ \hline
\multirow{5}{*}{\textbf{Anti-covidvax }  }     & United States (174,706), United Kingdom   (38,664), Canada (21,059), Australia (10,562), India (7,848), South Africa   (5,638), Ireland (3,893), New Zealand (1,985), Nigeria (928), Spain (907),   Malaysia (891), Republic of the Philippines (866), Germany (848), The   Netherlands (696), The Netherlands (696), The Netherlands (696), Jamaica   (595), France (649), Italy (586), Thailand (567), Mexico (551), Kenya (541) \\ \hline
\multirow{5}{*}{\textbf{Informative-covidvax}} & United States (64,203), India (19,852), United Kingdom (17,527), Canada (14,164), Australia (6,341), Ireland (2,734),   South Africa (2,376), Republic of the Philippines (2,045), Pakistan (1,116),   Nigeria (991), Kenya (948), Thailand (910), New Zealand (902), Malaysia   (866), Germany (753), Uganda (707), Spain (637), Italy (608), Sri Lanka   (568), France (527) \\ \hline
\multirow{5}{*}{\textbf{Hesitant-covidvax }}    & United States (209,109), United Kingdom   (54,740), Canada (28,016), India (26,315), Australia (15,477), South Africa   (9,528), Ireland (6,393), Republic of the Philippines (3,534), New Zealand   (3,035), Malaysia (2,967), Nigeria (2,095), Germany (1,844), The Netherlands   (1,782), Pakistan (1,642), Spain (1,605), Kenya (1,577), France (1,474), Italy   (1,150), Jamaica (959), Thailand (910)                                  \\ \hline
\end{tabular}
\end{table}

\begin{figure}
\centering
\includegraphics[width=0.8\textwidth]{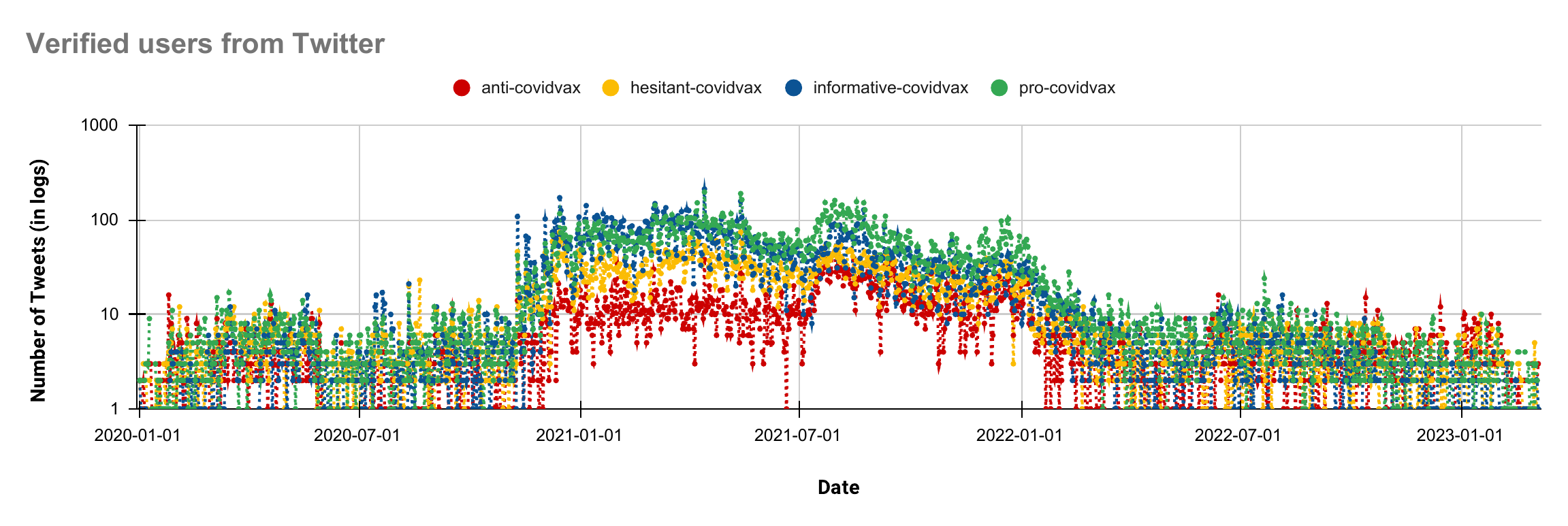}
\caption{The daily distributions of tweets for verified users across each class.}
\label{category wise verified users}
\end{figure}

In Figure \ref{category wise verified users}, we present the daily distribution of tweets created by accounts verified on Twitter. We observe that verified accounts post the highest number of Pro-covidvax and Informative-covidvax tweets, followed by Hesitant-covidvax and Anti-covidvax. This pattern could be due to the fact that verified users are often individuals of public interest or have significant influence in their respective fields, providing them with access to scientific literature, expert opinions, and public health guidance. Consequently, their views on COVID-19 vaccines and vaccination may be influenced, leading them to use Twitter as a platform to promote positive messages regarding COVID-19 vaccination.

\begin{figure}
\centering
\includegraphics[width=0.8\textwidth]{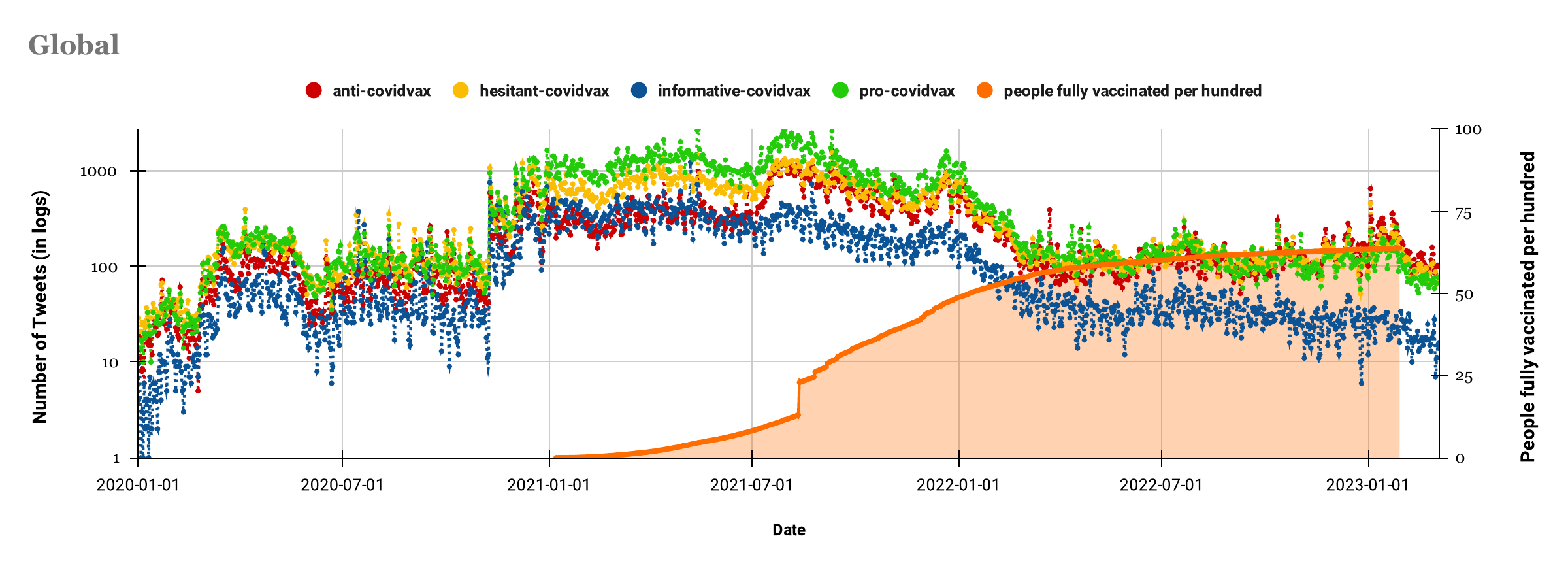}
\caption{The class-wise daily distribution of the tweets globally and the cumulative number of people vaccinated per hundred.}
\label{global daily-dist per category}
\end{figure}

Similarly, Figure \ref{global daily-dist per category} presents the class-wise daily distribution of tweets alongside the cumulative number of people fully vaccinated per hundred. The distribution reveals that prior to December 2020, individuals were skeptical of vaccines and vaccination, as tweets were a mix of Hesitant-covidvax, Informative-covidvax, Anti-covidvax, and Pro-covidvax. However, as the vaccination rollout progressed in December 2020 worldwide\footnote{https://ourworldindata.org/covid-vaccinations}, the trend shifted toward more Pro-covidvax and Informative-covidvax tweets regarding vaccines and vaccination, followed by  Hesitant-covidvax and Anti-covidvax tweets. This indicates a shift in the public's perception of COVID-19 vaccines and vaccination as more people become aware of their benefits.

The distribution of tweets among different classes was examined for the top six countries in the discourse. Figure \ref{three-graphs1} clearly demonstrates that the number of daily tweets related to vaccines and vaccination was lower prior to the start of vaccination. However, after the vaccine was rolled out at various times in different countries, the number of tweets in each class appeared to increase a few weeks earlier than the actual vaccination launch date. Once the vaccination process began, there was a significant surge in the tweets for all countries. Following this surge, the volume of tweets in the Pro-covidvax class was consistently higher than in the Anti-covidvax class. The Hesitant-covidvax class had the second-highest volume of tweets across all six countries after the Pro-covidvax class. Interestingly, after January 2022, the number of tweets in each class began to decline as 70\% of the population in each country had been vaccinated by then. However, this trend was not observed in South Africa, where the daily distribution of tweets is sparse and only over 35\% of the population is fully vaccinated as of March 2023.

\begin{figure}
     \centering
     \begin{subfigure}[b]{0.80\textwidth}
         \centering
         \includegraphics[width=1\textwidth]{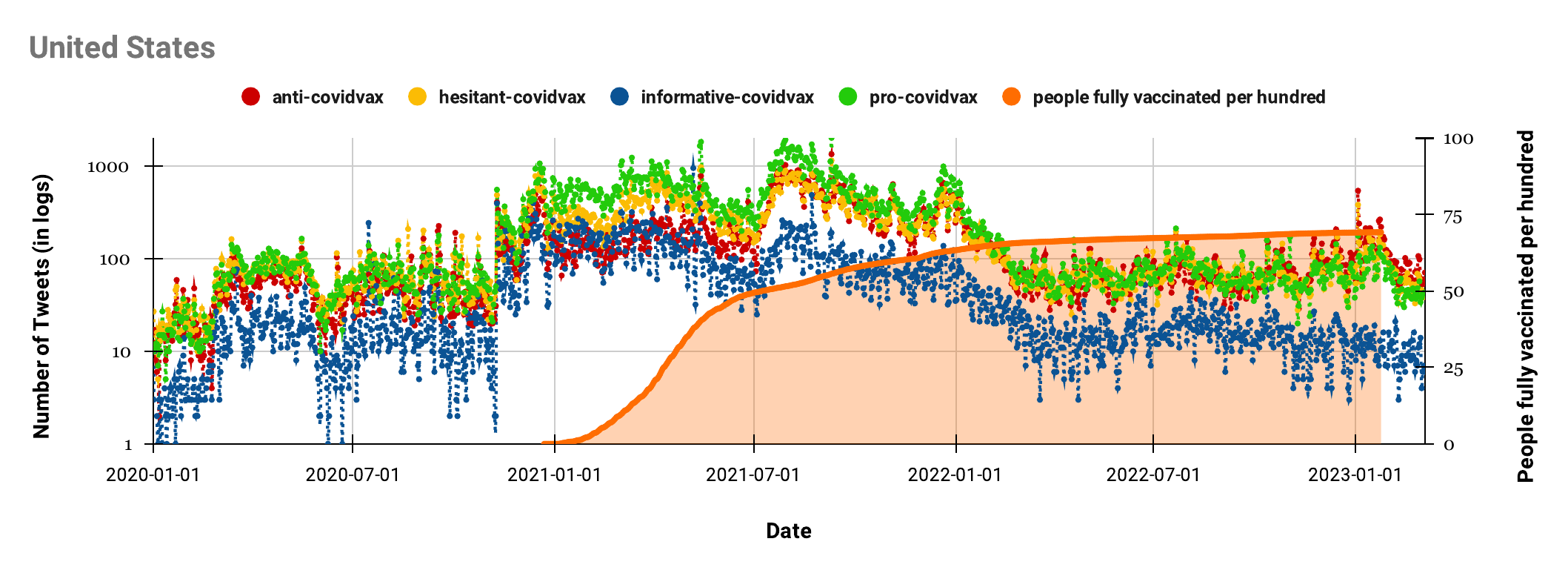}
        \caption{Four noticeable peaks of COVID-19 confirmed cases waves were observed in the country, i.e., 12-Jan-2021, 08-Sept-2021, 16-Jan-2022, and 28-July-2022.}
         \label{category-fig:1}
     \end{subfigure}
     \hfill
     \begin{subfigure}[b]{0.80\textwidth}
         \centering
         \includegraphics[width=1\textwidth]{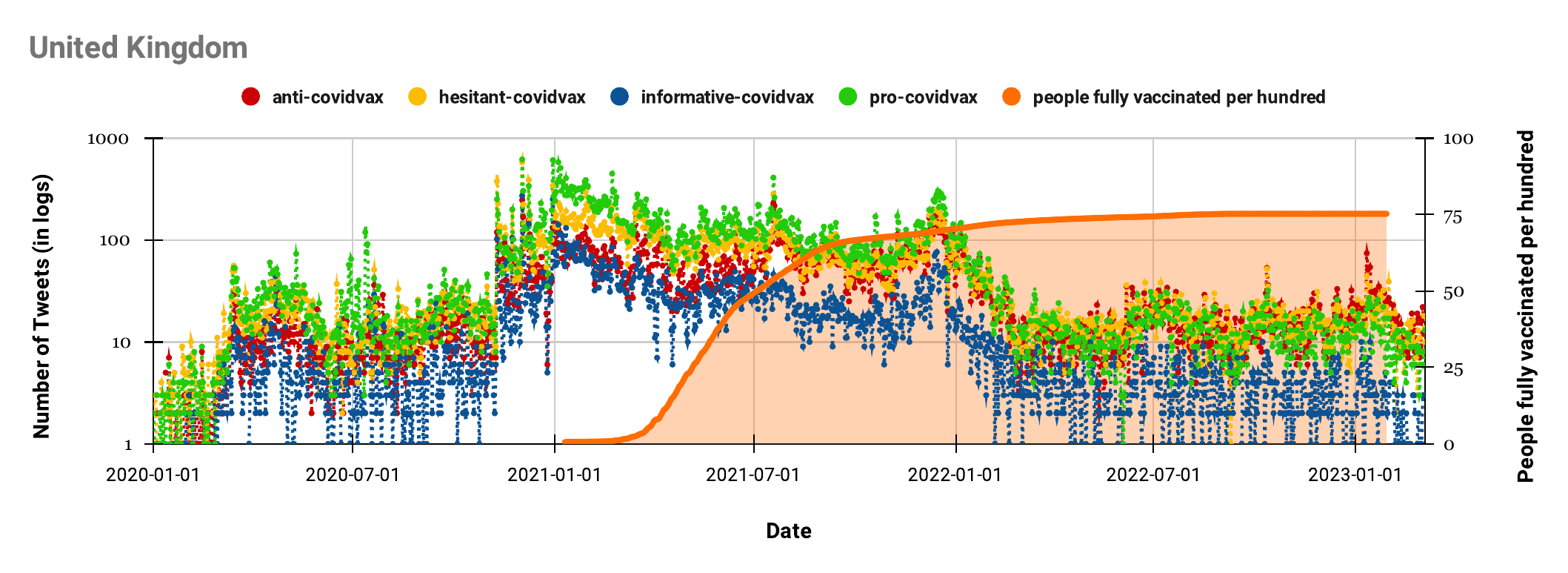} 
         \caption{Four noticeable peaks of COVID-19 confirmed cases waves were observed in the country, i.e., 07-Jan-2021, 21-July-2021, 07-Jan-2022, and 02-Mar-2022.}
         \label{category-fig:2}
     \end{subfigure}
     \begin{subfigure}[b]{0.80\textwidth}
         \centering
        \includegraphics[width=1\textwidth]{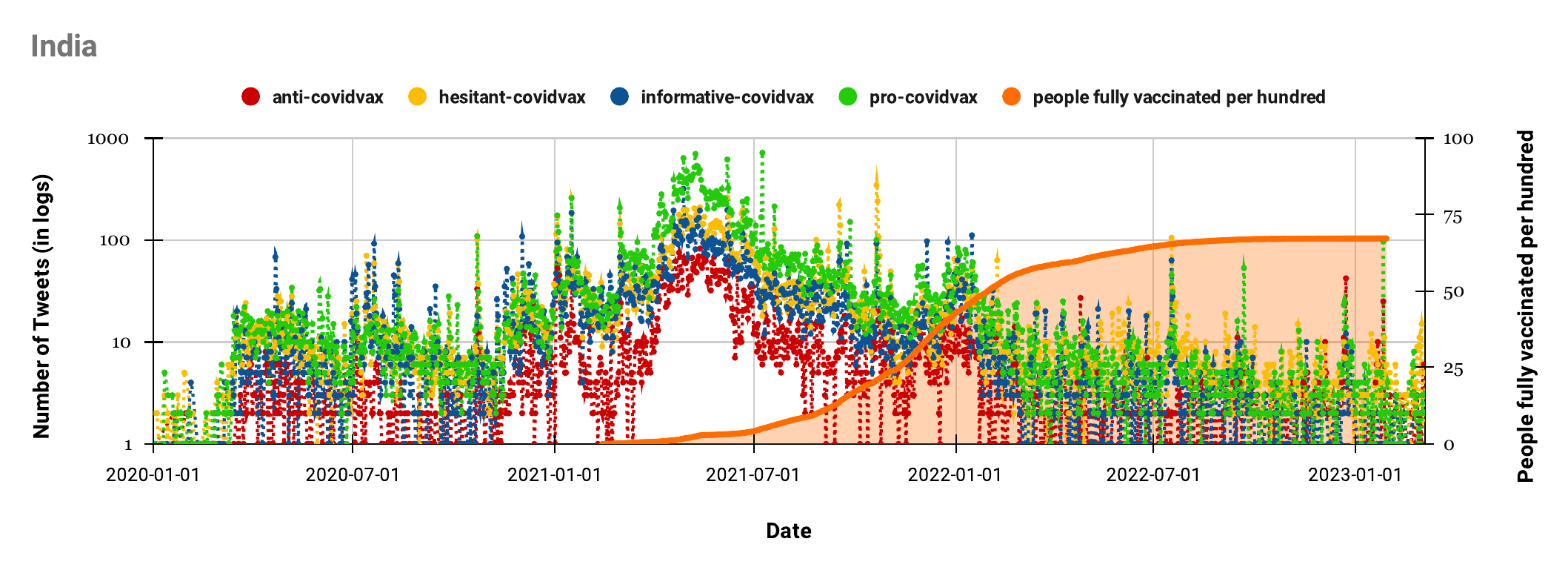}
        \caption{Three noticeable peaks of COVID-19 confirmed cases waves were observed in the country, i.e., 17-Sept-2020, 10-May-2021, and 25-Jan-2022.}
         \label{category-fig:3}
     \end{subfigure}  

\caption{Class-wise daily distributions of tweets in the United States, the United Kingdom, and India, and their cumulative vaccination data. (figure continued on the next page)}

\end{figure}

\begin{figure}\ContinuedFloat
\centering
     \begin{subfigure}[b]{0.80\textwidth}\ContinuedFloat
         \centering
         \includegraphics[width=1\textwidth]{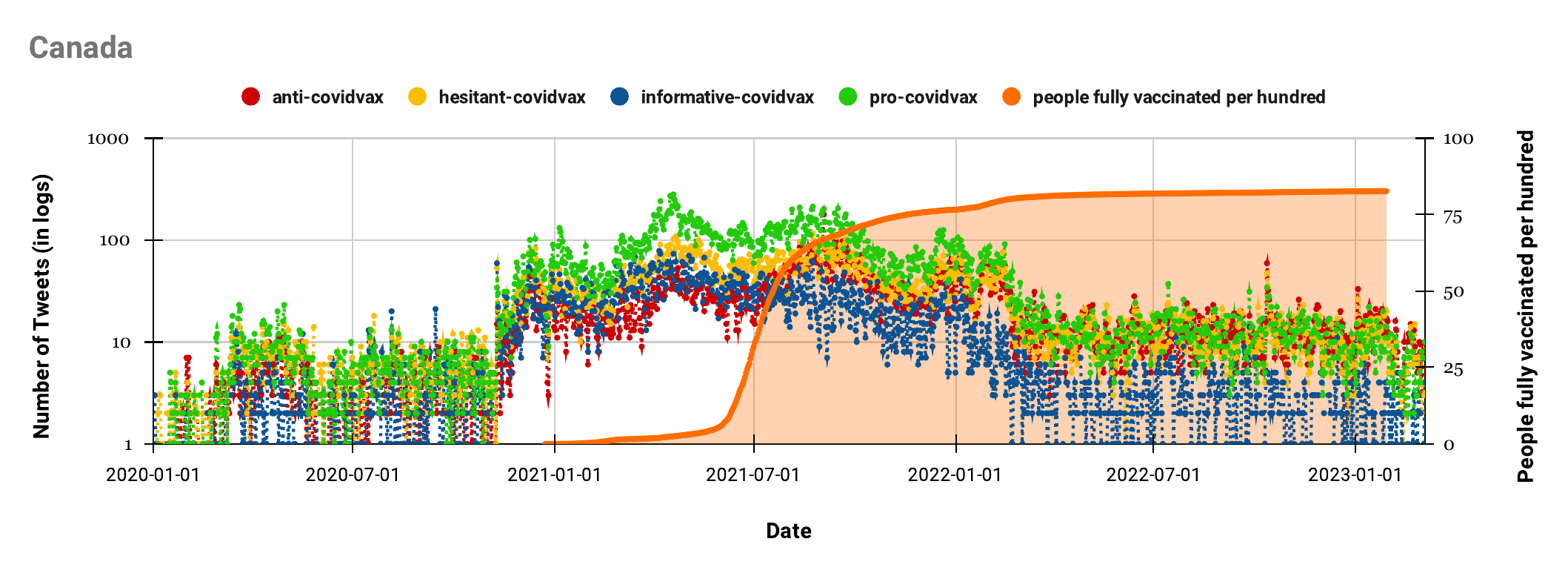}
         \caption{Six noticeable peaks of COVID-19 confirmed cases waves were observed in the country, i.e., 14-Jan-2021, 20-April-2021, 29-Sept-2021, 07-Jan-2022, 14-April-2022  and 01-Aug-2022.}
         \label{category-fig:4}
     \end{subfigure}  
     
     \begin{subfigure}[b]{0.80\textwidth}
         \centering
         \includegraphics[width=1\textwidth]{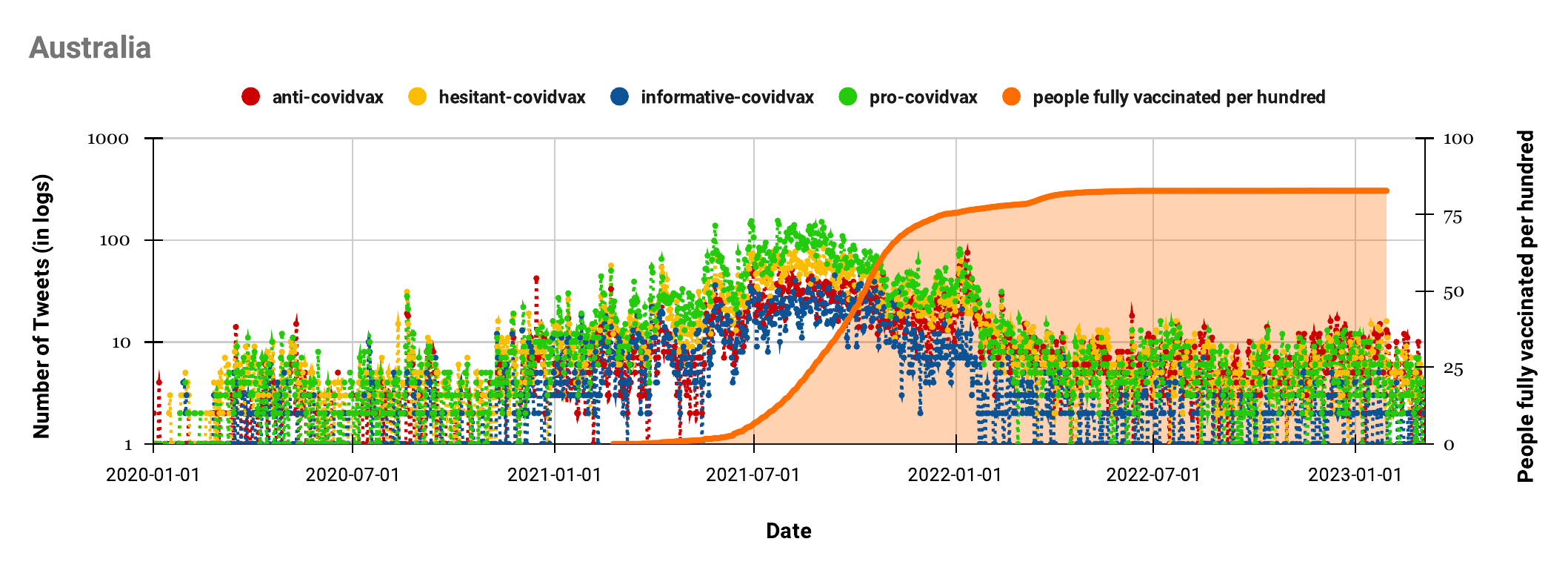}
         \caption{Five noticeable peaks of COVID-19 confirmed cases waves were observed in the country, i.e., 13-Jan-2022, 02-Feb-2022, 22-Mar-2022, 09-May-2022 and 23-July-2022.}
         \label{category-fig:5}
     \end{subfigure}
     
     \begin{subfigure}[b]{0.80\textwidth}
         \centering
         \includegraphics[width=1\textwidth]{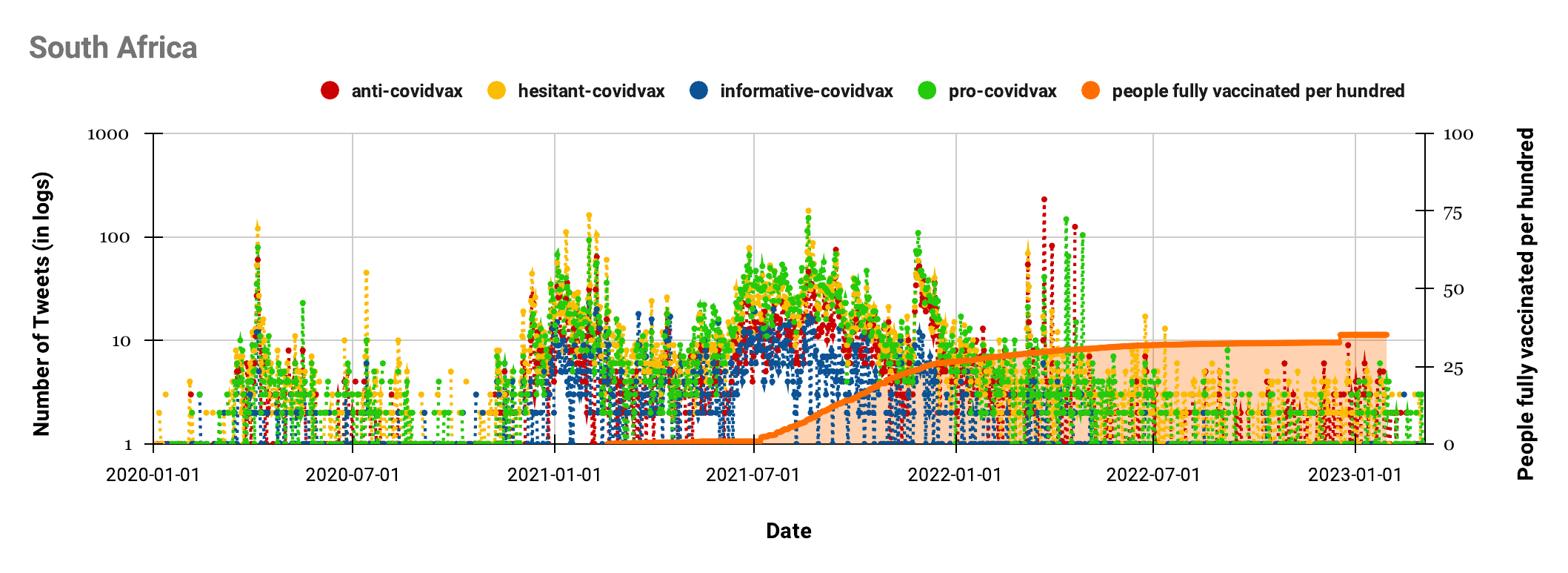}
         \caption{Five noticeable peaks of COVID-19 confirmed cases waves were observed in the country, i.e., 20-July-2020, 13-Jan-2021, 09-July-2021, 17-Dec-2021 and 15-May-2022.}
         \label{category-fig:6}
     \end{subfigure}

\caption{(Figure continued from the previous page) Class-wise daily distributions of tweets in Canada, Australia, and South Africa and their cumulative vaccination data.}
\label{three-graphs1}
\end{figure}

India followed a different trend for Anti-covidvax class compared to other countries, as the Anti-covidvax tweets were less the majority of the time throughout 2020--2023. The trend continued to be true even after the start of vaccinations in the country, indicating that most tweets from India were supportive, hesitant, or informative. This may be due to increased numbers of people who have received their vaccine doses and shared positive experiences on Twitter. A similar trend was reported in \citep{sv2021indian}. Similarly, in the case of South Africa, as shown in Figure \ref{category-fig:6}, we find the country's tweet distributions to be significantly sparse compared to other countries, as the discourse in the country around vaccines and vaccination seems to be insignificant on Twitter. This is also why we selected only the top six countries in the discourse for country-specific analysis.

We also examined the relationship between the volume of tweets and the COVID-19 waves that hit different countries at different times. In Figures \ref{category-fig:1}--\ref{category-fig:4}, we observe that after the vaccination rollout in the US, the UK, Canada, and India, there is a sharp decline in the volume of tweets across each class for a limited timeline as the number of COVID-19 confirmed cases per week peaked, except for Australia and South Africa. This trend continued until 70\% of the population was fully vaccinated per hundred.

\begin{figure}
     \centering
     \begin{subfigure}[b]{0.80\textwidth}
        \centering
        \includegraphics[width=0.8\textwidth]{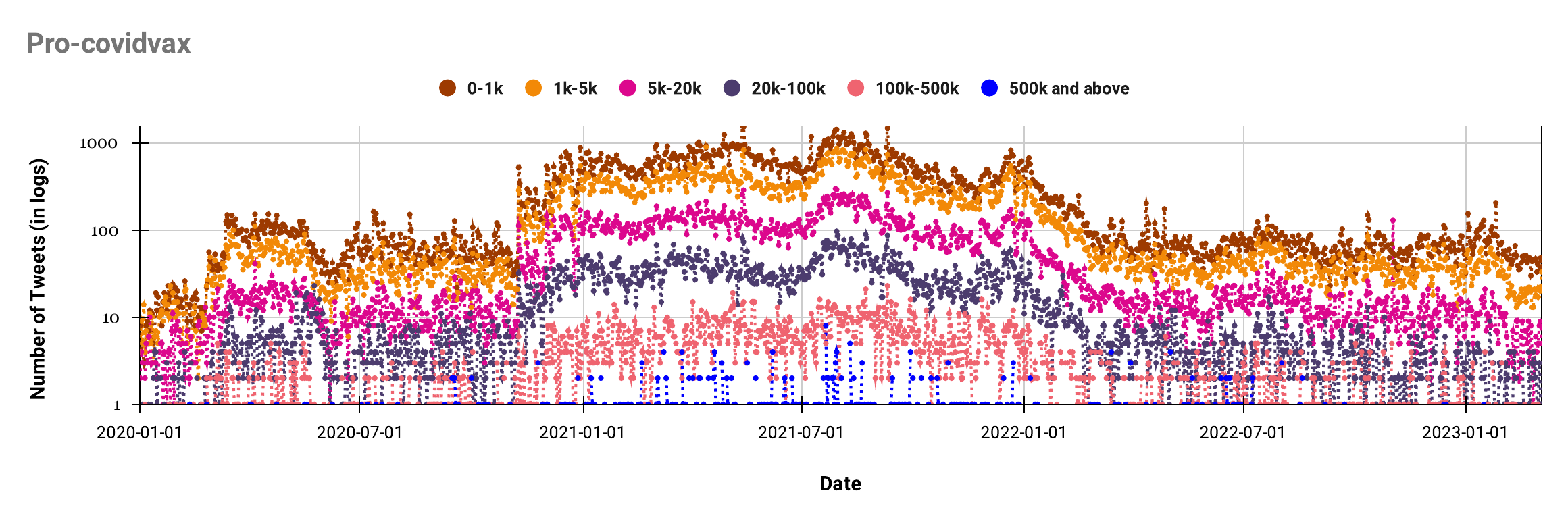}
        \caption{Pro-covidvax class}
        \label{daily-dist pro}
     \end{subfigure}
     \hfill
     \begin{subfigure}[b]{0.80\textwidth}
         \centering
         \includegraphics[width=0.8\textwidth]{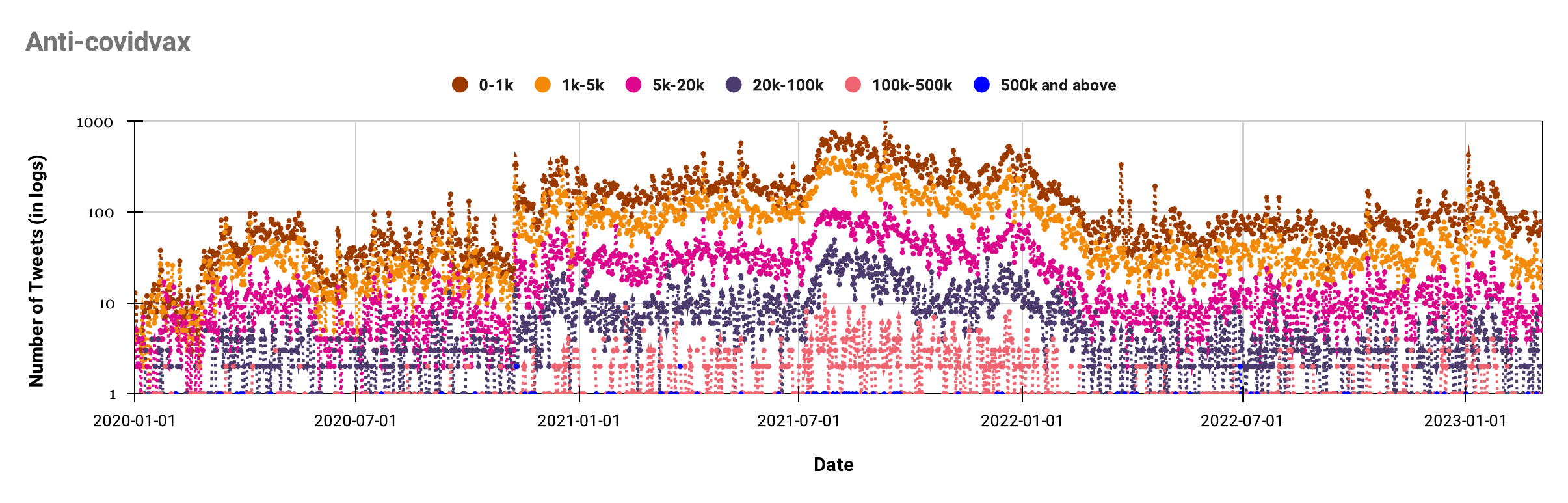}
         \caption{Anti-covidvax class}
         \label{daily-dist anti}
     \end{subfigure}
     \hfill
     \begin{subfigure}[b]{0.80\textwidth}
         \centering
        \includegraphics[width=0.8\textwidth]{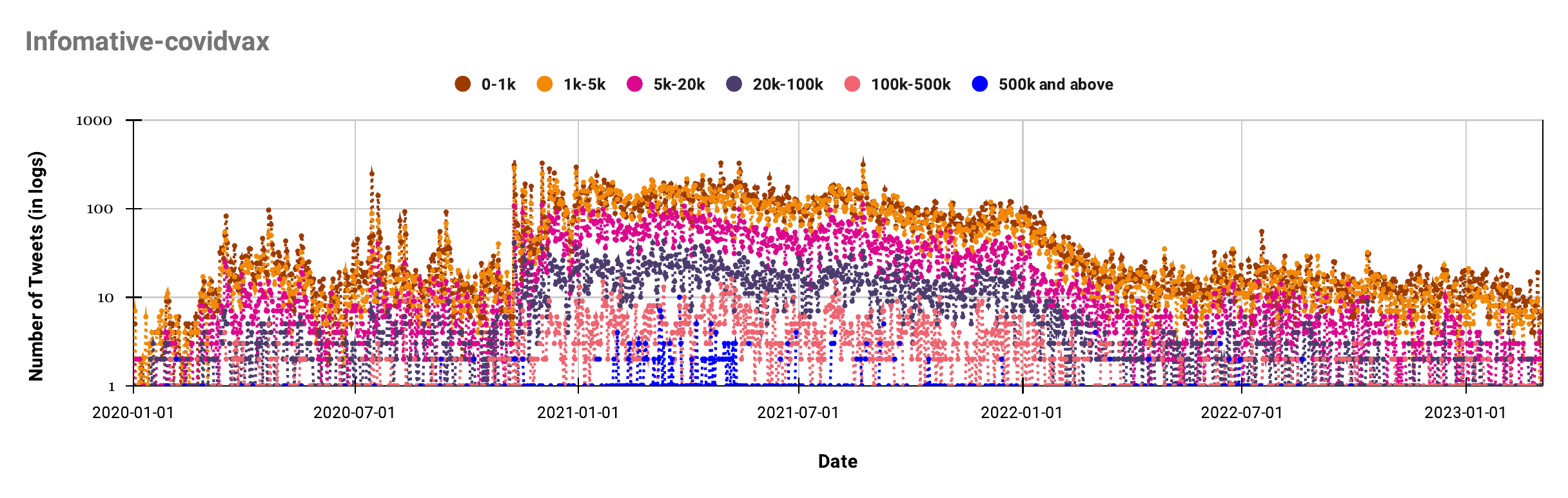}
        \caption{Informative-covidvax class}
        \label{daily-dist info}
     \end{subfigure}      
    \hfill
     \begin{subfigure}[b]{0.80\textwidth}
         \centering
        \includegraphics[width=0.8\textwidth]{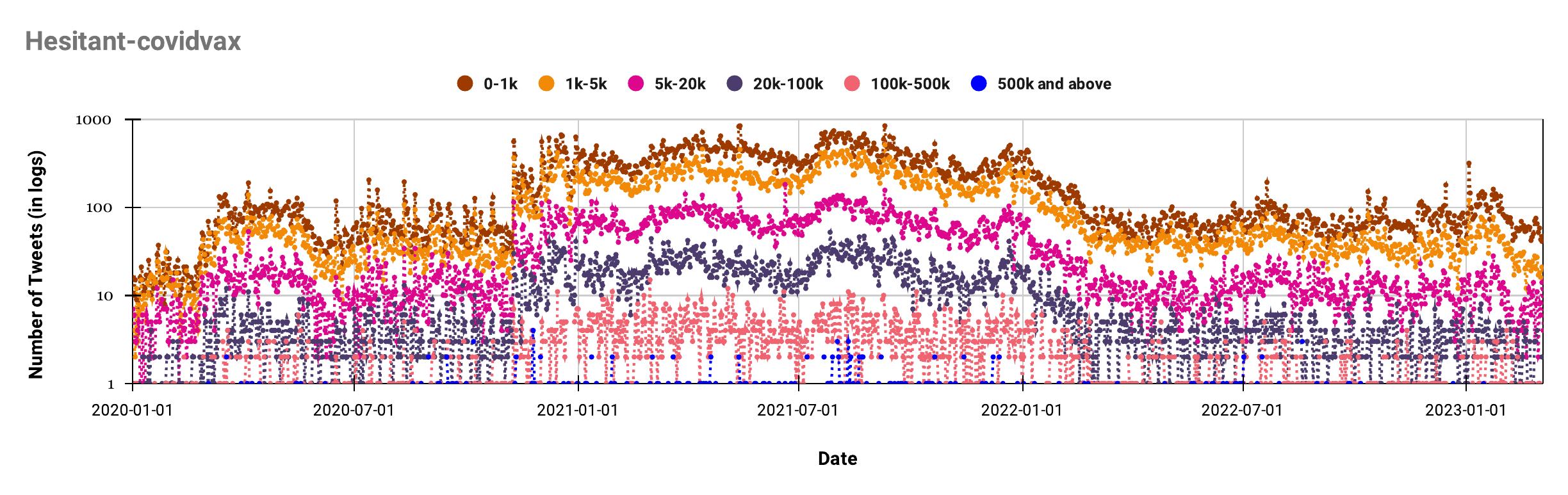}
        \caption{Hesitant-covidvax class}
        \label{daily-dist hesit}
     \end{subfigure} 
\caption{Class-wise distribution of tweets for each follower band.}
\label{four-graphs-followers}
\end{figure}

Further, we studied the distributions of tweets across classes versus the number of followers on Twitter. We divided the followers into six bands 0--1k, (1--5)k, (5--20)k, (20--100)k, (100--500)k, and 500k and above for four distinct vaccine classes: Pro-covidvax, Anti-covidvax, Informative-covidvax, and Hesitant-covidvax. As illustrated in Figure \ref{four-graphs-followers}, users with a follower count, such as 500k and above, tend to tweet less frequently. On the other hand, those with (0-1)k followers tend to post the highest number of tweets, followed by users in the (1-5)k, (5-20)k, (20-100)k, (100-500)k, and 500k and above bands for each class except Informative-covidvax (Figure \ref{daily-dist info}). In Informative-covidvax, we find equal dominance of 0-1k and (1-5)k followed by (5-20)k, (20-100)k, (100-500)k, and 500k and above. 

\begin{figure}
     \centering
     \begin{subfigure}[b]{0.80\textwidth}
        \centering
   \includegraphics[width=0.8\textwidth]{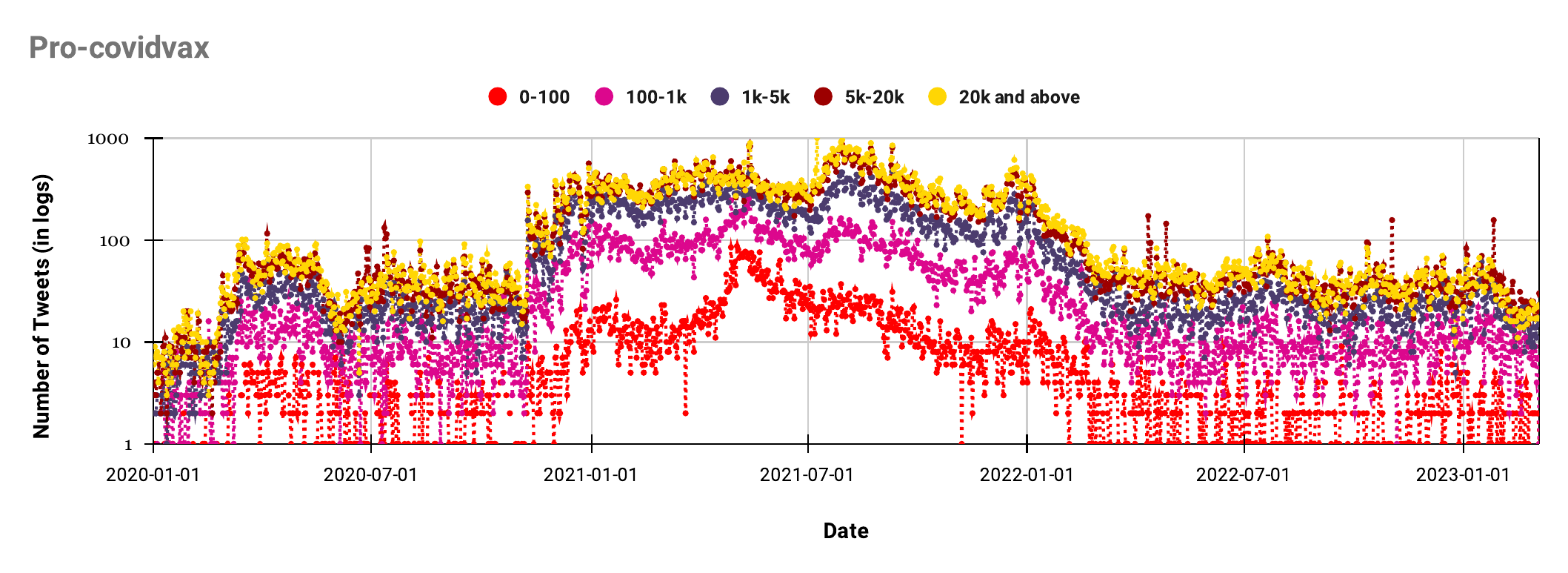}
        \caption{Pro-covidvax class}
       \label{daily-dist pro1}
     \end{subfigure}
     
     \begin{subfigure}[b]{0.80\textwidth}
         \centering
         \includegraphics[width=0.8\textwidth]{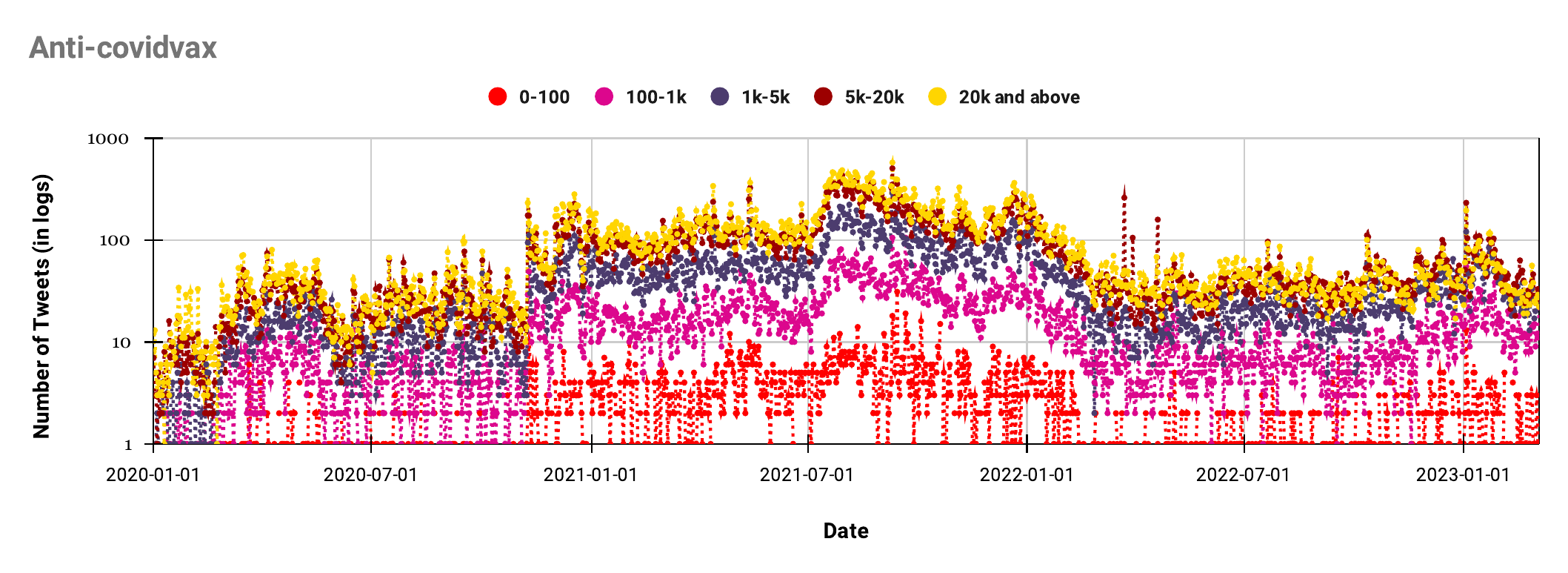}
         \caption{Anti-covidvax class}
         \label{daily-dist anti1}
     \end{subfigure}
     
     \begin{subfigure}[b]{0.80\textwidth}
         \centering
        \includegraphics[width=0.8\textwidth]{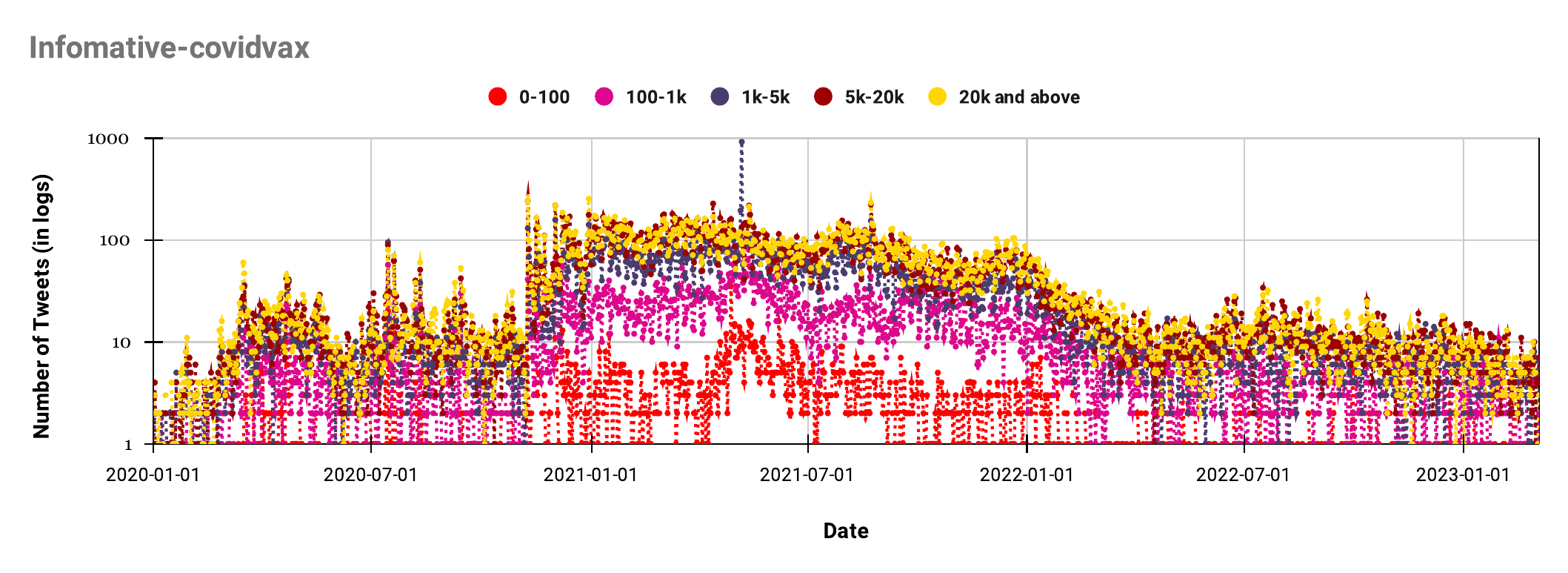}
        \caption{Informative-covidvax class}
        \label{daily-dist info1}
     \end{subfigure}      
    
     \begin{subfigure}[b]{0.80\textwidth}
         \centering
        \includegraphics[width=0.8\textwidth]{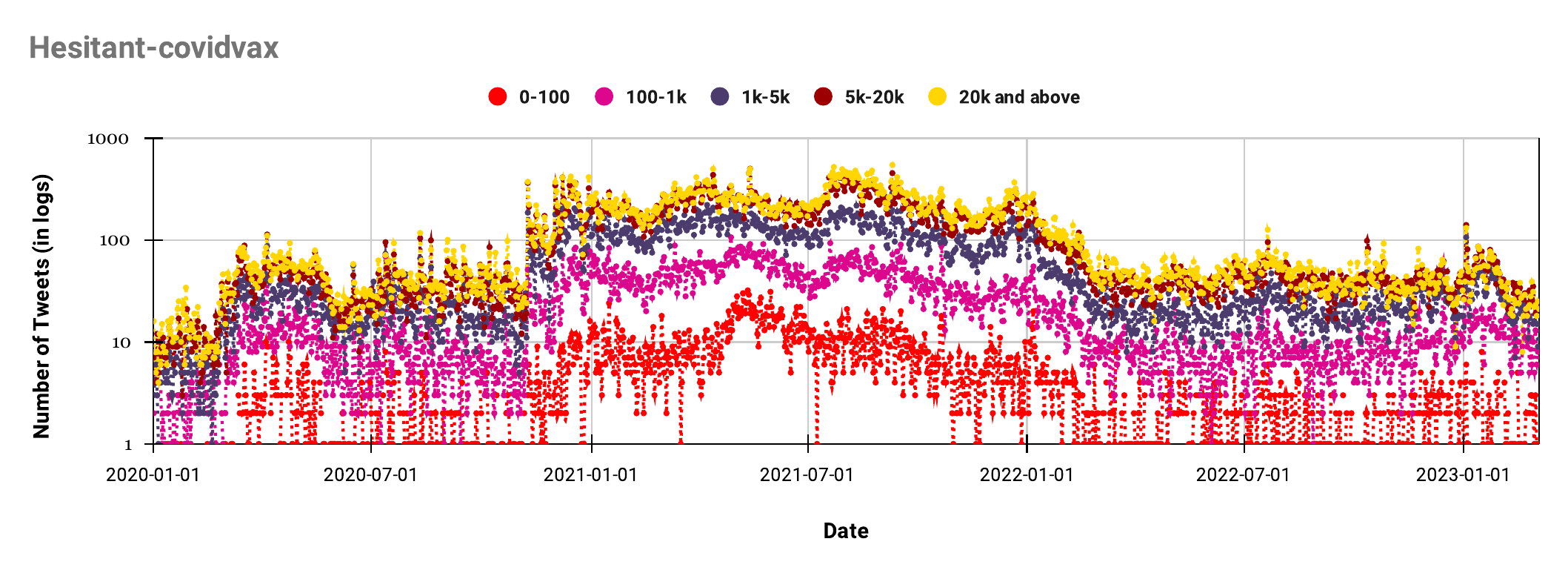}
        
        \caption{Hesitant-covidvax class}
        \label{daily-dist hesit1}
     \end{subfigure} 
\caption{Class-wise distribution of tweets for each tweet count band.}
\label{four-graphs-count}
\end{figure}

There is an interesting trend among Twitter users with 500k or above follower count following the vaccination rollout. As depicted in Figure \ref{four-graphs-followers}, that group seemed to be sharing Informative-covidvax tweets after the vaccination rollout and until the next couple of weeks, compared to other timelines. The figure also indicates that compared to Pro-covidvax and Hesitant-covidvax, these influential users are more inclined to promote the vaccine and its benefits. The absence of Anti-vaccination tweets by these users, as seen in Figure \ref{daily-dist anti}, suggests that they endorse the COVID-19 vaccine and advocate for its widespread adoption. Overall, the findings suggest a positive attitude of influential Twitter accounts towards COVID-19 vaccination. Furthermore, we analyzed the relationship between the tweet counts of Twitter users and their tweeting behaviours in the four classes. We categorized tweet counts into five bands to conduct the analysis: 0-100, 100-1k, 1k-5k, 5k-20k, and 20k and above. As shown in Figure \ref{four-graphs-count}, users with a tweet count of 20k and above tend to post the highest number of tweets across each class, followed by those with a count of 5k-20k, 1k-5k, 100-1k, and 0-100. Also, Pro-covidvax tweets are always higher than other classes during January 2021--January 2022, when most countries significantly pushed their fully vaccinated per hundred numbers.

Also, we analyzed each class' top 5 tweet interests for the overall timeline against followers and tweet count contexts. In the case of followers, the 0-1k band seemed to dominate the discourse across all classes. The Pro-covidvax and Hesitant-covidvax attained their top tweet interests from the 0-1k band in May 2021, July 2021, and September 2021. For Anti-covidvax, the highest tweet interest was observed during July 2021, August 2021, and September 2021. In the case of Informative-covidvax, the band pushed the tweet interests significantly in November 2020, December 2020, April 2021, May 2021, and August 2021. Similarly, in the case of tweet count, the 20k and above band seemed to dominate the discourse across all classes, followed by 5k--20k on three occasions and 1k--5k on a single occasion. In Pro-covidvax class, all top 5 tweet interests were generated by the 20k and above band in May 2021, July 2021, and August 2021. In Anti-covidvax class, the 20k and above band generated the highest tweet interest on four occasions in July--September 2021, with the 5k-20k band generating the second highest tweet interest in September 2021. Similarly, the Hesitant-covidvax class received the highest tweet interest from the 5k-20k band in April 2021, May 2021, July 2021, and September 2021, with the 5k-20k band generating the third highest tweet interest in May 2021. Interestingly, in the case of Informative-covidvax class, the highest tweet interest was generated by the 1k-5k band in May 2021, with the 5k-20k band in the third during November 2020 and the 20k and above band taking up the other positions.

\subsection{Theoretical and practical implications}
In this study, we analyzed the discourse surrounding COVID-19 vaccines and vaccination on Twitter in response to the ongoing pandemic and the need to better understand the global conversation through numerous spatiotemporal analyses. The existing research varied in terms of regional focus, languages considered, the methods employed for stance and sentiment analysis, and the level of fine-grain classification of stances across multiple conversational dynamics \citep{deverna2021covaxxy, muric2021covid, chen2022multilingual, di2022vaccineu, hu2021revealing, kwok2021tweet, khan2021us,luo2022understanding,qorib2023covid}. To expand the research in the area, we curated and analyzed a large-scale geotagged vaccines and vaccination-related Twitter dataset, classified tweets as pro, anti, informative, hesitant, or not-related and exploring the conversational dynamics across global and country-specific engagements, notable accounts, vaccination progress, and contextual hashtags, to name a few.

The findings of this social media analysis on vaccines and vaccination related to COVID-19 have several practical and theoretical implications. Practically, the study can help public health officials and policymakers design effective communication strategies to address vaccine hesitancy and promote vaccination. By analyzing the distribution of tweets across different classes and geographical regions, policymakers can identify the key concerns and barriers people have towards vaccination. For instance, the study highlights the presence of negative hashtags such as \#vaccinesideeffects and \#plandemic in the Anti-covidvax class, indicating that people who are hesitant about vaccines are concerned about the safety and efficacy of vaccines. On the other hand, positive hashtags such as \#getvaccinated, \#vaccineswork, and \#vaccinessavelives in the Pro-covidvax class indicate that people who support vaccines are motivated by the benefits of vaccination. Based on these insights, policymakers can tailor their communication strategies to address different groups' specific concerns and motivations, thereby improving vaccine uptake and acceptance.

The study also has theoretical implications. By analyzing the daily distribution of tweets across different classes, the study sheds light on the dynamics of public opinion and discourse on vaccines and vaccination. The shift towards more Pro-covidvax and Informative-covidvax tweets regarding vaccines and vaccination after the vaccination rollout suggests that people's perceptions of vaccines and vaccination are dynamic and can change over time. This finding is consistent with the theoretical framework of the Diffusion of Innovation theory \citep{rogers2010diffusion}, which suggests that people's adoption of new ideas, behaviours, or technologies is influenced by social factors, including the perceived benefits and risks, social norms, and communication channels. The study also highlights the importance of social influence in shaping public opinion on vaccines and vaccination, as verified accounts post the highest number of Pro-covidvax and Informative-covidvax tweets. This finding is consistent with the social identity theory, which suggests that people's opinions and behaviours are influenced by their groups and social norms.

In summary, the practical and theoretical implications of this social media analysis suggest that policymakers and public health officials can design effective communication strategies by tailoring their messages to address different groups' specific concerns and motivations. The study also highlights the importance of social influence in shaping public opinion and suggests that people's perceptions of vaccines and vaccination are dynamic and can change over time.

\section{Conclusion}
This paper presents an infoveillance study of COVID-19 vaccines and vaccination discourse on Twitter. First, the paper introduced \textit{GeoCovaxTweets Extended}, a comprehensive and multilingual Twitter dataset that contains 2.8 million tweets related to COVID-19 vaccines and vaccination discourse. The dataset contains tweets created by over 650k users from 238 countries and territories from January 01, 2020, to March 05, 2023. Second, a labelled dataset called \textit{PAIHcovax} was developed, including the Pro, Anti, Informative, and Hesitant contexts surrounding COVID-19 vaccines and vaccination discourse. Third, multiple transformer-based models were experimented on the \textit{PAIHcovax} dataset to design a classifier, \textit{CovaxBERT}. Using this classifier on \textit{GeoCovaxTweets Extended}, we extensively examined the spatial-temporal trends of each stance and the dominant themes in the discourse. Our study provides insights into the global discussions around COVID-19 vaccines, which can be useful for policymakers and health practitioners in their efforts to increase vaccine uptake and reduce hesitancy for future pandemics or epidemics.

This study has multiple limitations, even after using Twitter's Full-archive search endpoint for data collection. Twitter users who engage in vaccine discourse may be different from the general population, which could limit the generalizability of findings. Additionally, we recognize that subjectivity can influence annotations, leading to biases. While we have provided clear instructions to annotators, we acknowledge that some level of subjectivity may still persist. Furthermore, specific geographic regions or demographic groups may be overrepresented in the data, limiting generalizability. Twitter users may delete or modify their tweets after they are posted, leading to missing or incomplete data. Language barriers may limit Infodemiology studies based on Twitter data. Twitter users can communicate in languages other than English, even in English-speaking nations, limiting the ability to analyze data in other languages. Twitter users' demographic data are often incomplete or unavailable, which is a limit of this study as generalizations to different populations are absent. With detailed demographic data, researchers can accurately assess how different groups engage in the vaccines and vaccination Twitter discourse. This is particularly important since the effectiveness of vaccine campaigns and messaging can depend on factors such as age, gender, or socioeconomic status.

\section*{CRediT authorship contribution statement}
\textbf{PS}, \textbf{RL}, and \textbf{M} performed Conceptualization, Data curation, Methodology, Software,
Visualization, Writing–first draft. \textbf{SC} performed Supervision, Writing–Review \& Editing. \textbf{BS} performed Software,
Visualization.

\section*{Declaration of Competing Interest}
The authors declare that they have no known competing financial interests or personal relationships that could have appeared to influence the work
reported in this paper.

\section*{Data availability}
Both the datasets --- \textit{GeoCovaxTweets Extended} and \textit{PAIHcovax} --- will be released from Harvard Dataverse. And, \textit{CovaxBERT} will be shared from Hugging Face. \textbf{The releases will be made after the peer review completes.}

\section*{Acknowledgements}
We (the authors) thank the \textit{Nectar Research Cloud} for providing us with a large-volume compute instance.

\bibliography{cas-refs.bib}% common bib file
%\bibliographystyle{apalike}
\appendix
\section{Some examples from the \textit{not-related} class.}
\begin{itemize}
\item Sadly some folks need to hit a wall. Salk vaccine used monkey kidney parts... they later found it causes cancer. Read Dr. MARY'S MONKEY. THE BOOK. 
\item Wishing my good friend Natalie Hart success this weekend. As President of the Provincial Liberal Party, she will rebuild and rejuvenate the party. Donating endless hours, drawing reluctant Malton residents to COVID vaccination clinics.
\item An important announcement today to create 5 hubs for vaccine and therapeutic research/production. A product of advocacy from Canada research universities ahead of Budget 2021 and a federal government willing to double down on areas of strength.
\item The greatest perpetrator of misinformation during the pandemic, has been the United States government the conspiracy theorist knew the truth,   all along.
\item Ads for a vaccine are crazy.   
\item Hey Tom Cotton as a good republican you are still attached to the lie. Was it the Democrats or was it Trump with his cohort of bootlicking Republicans who played down the warnings from the Chinese?  And even today they deny the value.      
\item It's awful, buring pain like your nerve endings are on fire, AND you can get in again\#Vaccinate against shingles if you had chicken pox.
\item Wishing my good friend Natalie Hart success this weekend. As President of the Provincial Liberal Party she will rebuild and rejuvenate the party. Donating endless hours, drawing reluctant Malton residents to COVID vaccination clinics, saved lives. She has strength, smarts...
\item Republicans in Congress seems commercial mindset on building economy but you have proven decrease deficit \& collected funds to fight for climate change, created employment, protected American citizens from covid-19 through vaccination camps.      
\item Millie literally wore me tf out taking her to get a simple nasal spray vaccine. Training can’t start soon enough. 

\end{itemize}

\end{document}